\documentclass[10pt]{article}

\usepackage{fullpage}
\usepackage[T1]{fontenc}
\usepackage[latin1]{inputenc}
\usepackage{url}
\usepackage{amsmath}
\usepackage{amssymb}
\usepackage{amsfonts}
\usepackage{stmaryrd}
\usepackage{enumerate}
\usepackage{txfonts}
\usepackage{pxfonts}
\usepackage{epsfig}
\usepackage{color}
\usepackage{array}
\usepackage{bussproofs}
\usepackage{xypic}
\xyoption{all}
\usepackage{natbib}

\newcommand{\todo}{\textbf{TODO: }}
\newcommand{\mc}{\mathcal}
\newcommand{\restr}{\hspace{-0.1cm}\upharpoonright}
\newcommand{\fl}{\to}

\newcommand{\eq}{\simeq_\varepsilon}
\newcommand{\reec}{\Rightarrow}
\newcommand{\cqfd}{\hfill\ensuremath{\Box}}

\newcommand{\bec}{\begin{center}}
\newcommand{\eec}{\end{center}}
\newcommand{\btab}{\begin{tabular}}
\newcommand{\etab}{\end{tabular}}

\newcommand{\AAA}{\mc{A}}
\newcommand{\LL}{\llbracket}
\newcommand{\RR}{\rrbracket}
\newcommand{\trad}[1]{\LL#1\RR}
\newcommand{\pair}[2]{\langle#1,#2\rangle}
\newcommand{\view}[1]{\ulcorner#1\urcorner}
\newcommand{\LLL}{\mc{L}}

\newcommand{\DD}{\mc{D}}
\newcommand{\FF}{\mc{F}}
\newcommand{\OO}{\mathbf{O}}
\newcommand{\PP}{\mathbf{P}}
\newcommand{\RRR}{\mc{R}}
\newcommand{\id}{\textit{id}}

\newcommand{\ie}{i.e.}

\newcommand{\negocc}{\textit{Neg}}
\newcommand{\posocc}{\textit{Pos}}

\newcommand{\occ}{\mathcal{O}}

\newcommand{\rref}{\textit{ref}}
\newcommand{\MM}{\mathcal{M}}
\newcommand{\EEE}{\mathcal{E}}
\newcommand{\erase}{\textit{erase}}
\newcommand{\dna}{{\mathop{\downarrow}}}
\newcommand{\upa}{{\mathop{\uparrow}}}
\newcommand{\paux}{\textit{paux}}
\newcommand{\orig}{\textit{or}}
\newcommand{\proof}{\textsc{Proof: }}

\newcommand{\tilsig}{\tilde{\sigma}}
\newcommand{\tiltau}{\tilde{\tau}}
\newcommand{\barsig}{\bar{\sigma}}
\newcommand{\bartau}{\bar{\tau}}
\newcommand{\barrho}{\bar{\rho}}
\newcommand{\friends}{\textit{fr}}

\newcommand{\restrun}{\restr_{\dna\upa,\dna\dna}}
\newcommand{\restrdeux}{\restr_{\upa,\dna\upa}}
\newcommand{\restrtrois}{\restr_{\upa,\dna\dna}}

\title{Curry-style type isomorphisms and game semantics}
\author{Joachim de Lataillade\\Preuves Programmes Systèmes\\CNRS - Paris 7\\Joachim.de-Lataillade@pps.jussieu.fr}
\date{}

\makeatletter
\renewcommand\paragraph{\@startsection{paragraph}{4}{\z@}%
                                      {1.25ex \@plus.5ex \@minus.2ex}%
                                      {-1em}%
                                      {\normalfont\normalsize\bfseries}}

\makeatother

\newcount\xxanglea
\newcount\xxangleb
\makeatletter
\def\anglea{\@arabic\xxanglea}
\def\angleb{\@arabic\xxangleb}
\makeatother
\newcommand\boucle[2][120]{
\xxanglea=#1
\xxangleb=\xxanglea
\advance\xxangleb by -60
\nccurve[angleA=\anglea,angleB=\angleb,ncurv=6,nodesep=0.15]{->}{#2}{#2}}%

\newenvironment{minilist}{\begin{list}{$\bullet$}{
  \setlength{\parsep}{0pt}
  \setlength{\topsep}{0pt}
  \setlength{\itemsep}{-\parsep}
  \setlength{\labelsep}{0.5em}}}{\end{list}}

\begin{document}

\newtheorem{Def}{Definition}
\newtheorem{Th}{Theorem}
\newtheorem{Ex}{Example}
\newtheorem{Prop}{Proposition}
\newtheorem{Lemme}{Lemma}
\newtheorem{Corollaire}{Corollary}
\newtheorem{Claim}{Claim}

\maketitle

\begin{abstract}
Curry-style system F, \ie\ system F with no explicit types in terms, can be
seen as a core presentation of polymorphism from the point of view of
programming languages.

\medskip

This paper gives a characterisation of type isomorphisms for this
language, by using a game model whose intuition comes both from the
syntax and from the game semantics universe. The model is composed of:
an untyped part to interpret terms, a notion of game to interpret
types, and a typed part to express the fact that an untyped strategy
$\sigma$ plays on a game $A$.

\medskip

By analysing isomorphisms in the model, we prove that the equational
system corresponding to type isomorphisms for Curry-style system F is
the extension of the equational system for Church-style isomorphisms
with a new, non-trivial equation: $\forall X.A\eq A[\forall Y.Y/X]$ if
$X$ appears only positively in $A$.
\end{abstract}

\section{Introduction}

\paragraph{Types isomorphisms.}The problem of type isomorphisms is a purely syntactical question:
two types $A$ and $B$ are isomorphic if there exist two terms $f:A\fl
B$ and $g:B\fl A$ such that $f\circ g=id_B$ and $g\circ f=id_A$. This
equivalence relation on data types allows to translate a program from
one type to the other without any change on the calculatory meaning of
the program. Thus, a search in a library up to type isomorphism will
help the programmer to find all the functions that can potentially
serve his purpose, and to reuse them in the new typing
context~\cite{rittri}. This is particularly appealing with functional
languages, because in this case the type can really be seen as a
partial specification of the program: such a library search up to
isomorphisms has been implemented in particular for Caml Light by
Jérôme Vouillon. It can also be used in proof assistants to help
finding proofs in libraries and reusing them~\cite{isoreusedep} (for
more details on the use of type isomorphisms in computer science,
see~\cite{isotypes}).  From a more general point of view, type
isomorphisms are the natural answer to the question of equivalence
between types in a programming language.

The question of characterising these type isomorphisms is then a very
simple problem to formulate, however its resolution is often
non-trivial, especially when dealing with polymorphism.  Roberto Di
Cosmo~\cite{isotypes} has solved syntactically this question for
\textit{Church-style} system F (\ie\ system F where types appear
explicitly in the terms) by giving an equational system on types
equivalent to type isomorphisms.  In a preceding work~\cite{lataillade}, we
have given a new proof of this result by using a game semantics model
of Church-style system F.  In this more geometrical approach, types were
interpreted by an arborescent structure, \textit{hyperforests}: the
natural equality for this structure happened to be exactly the equality
induced by type isomorphisms.  The efficiency of game semantics in
this context was an incitement to go further and to explore the
possibility of resolving this question for other languages.

\begin{figure}[h!]
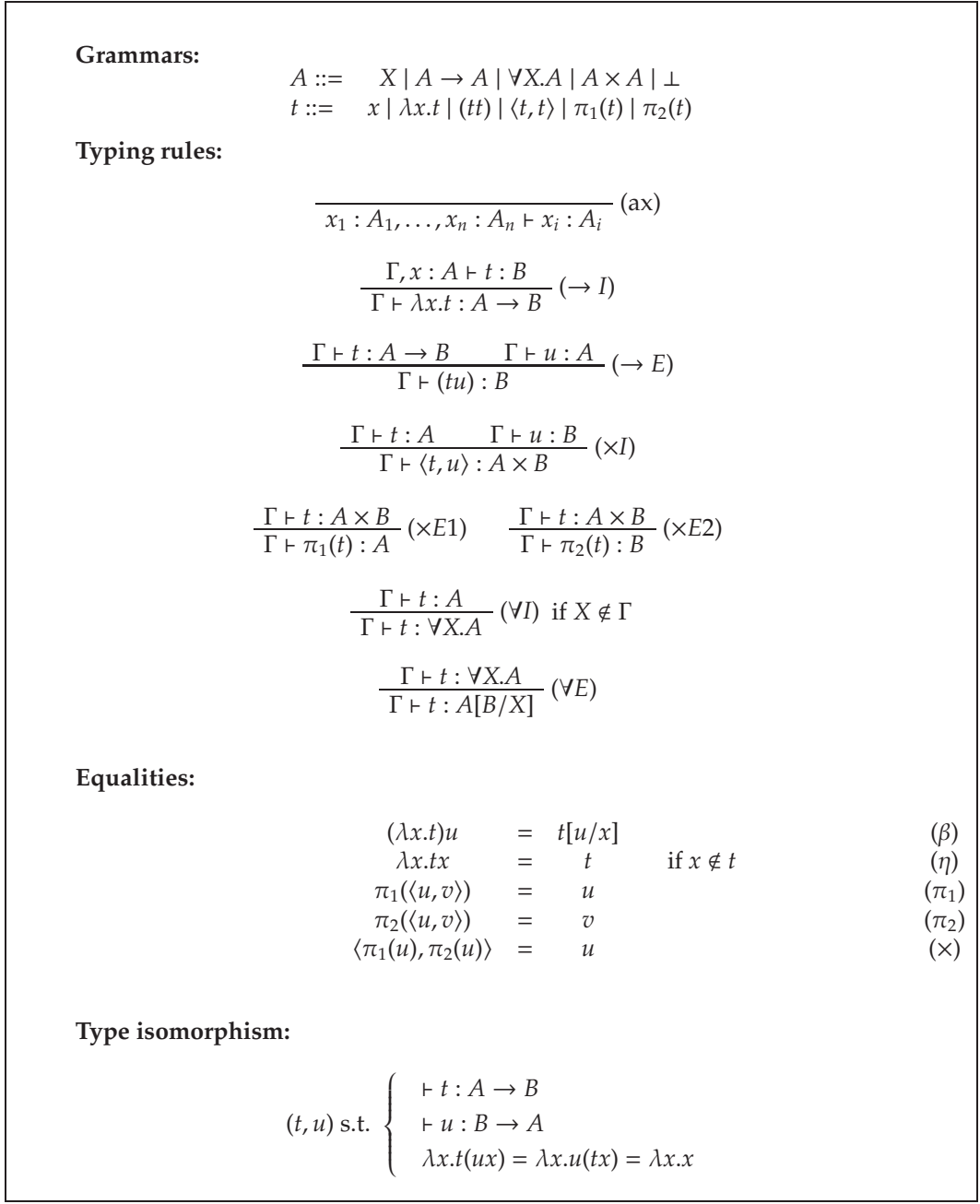
\label{systF}
\bec
\begin{tabular}{|@{\hspace{1cm}}c@{\hspace{1cm}}|}\hline\\
\bigskip
\parbox[c]{12cm}{
\paragraph{Grammars:}

$$
\begin{array}{lcr}
A::=\quad&X\mid
A\fl A\mid\forall X.A\mid A\times A\mid\bot\\
t::=\quad&x\mid \lambda x.t\mid (tt)\mid\langle t,t\rangle\mid\pi_1(t)\mid\pi_2(t)\\
\end{array}
$$

\paragraph{Typing rules:}

\bec
\AxiomC{\phantom{aa}}
\RightLabel{(ax)}
\UnaryInfC{$x_1:A_1,\dots,x_n:A_n\vdash x_i:A_i$}
\DisplayProof

\bigskip

\AxiomC{$\Gamma,x:A\vdash t:B$}
\RightLabel{($\rightarrow I$)}
\UnaryInfC{$\Gamma\vdash\lambda x.t:A\rightarrow B$}
\DisplayProof

\bigskip

\AxiomC{$\Gamma\vdash t:A\rightarrow B$}
\AxiomC{$\Gamma\vdash u:A$}
\RightLabel{($\rightarrow E$)}
\BinaryInfC{$\Gamma\vdash (tu):B$}
\DisplayProof

\bigskip

\AxiomC{$\Gamma\vdash t:A$}
\AxiomC{$\Gamma\vdash u:B$}
\RightLabel{($\times I$)}
\BinaryInfC{$\Gamma\vdash \pair{t}{u}:A\times B$}
\DisplayProof

\bigskip

\AxiomC{$\Gamma\vdash t:A\times B$}
\RightLabel{($\times E1$)}
\UnaryInfC{$\Gamma\vdash \pi_1(t):A$}
\DisplayProof
\quad\AxiomC{$\Gamma\vdash t:A\times B$}
\RightLabel{($\times E2$)}
\UnaryInfC{$\Gamma\vdash \pi_2(t):B$}
\DisplayProof

\bigskip

\AxiomC{$\Gamma\vdash t:A$}
\RightLabel{($\forall I$)}
\UnaryInfC{$\Gamma\vdash t:\forall X.A$}
\DisplayProof if $X\notin\Gamma$

\bigskip

\AxiomC{$\Gamma\vdash t:\forall X.A$}
\RightLabel{($\forall E$)}
\UnaryInfC{$\Gamma\vdash t:A[B/X]$}
\DisplayProof

\bigskip

\eec

\paragraph{Equalities:}

$$
\begin{array}{@{\hspace{4cm}}cccc@{\hspace{2.7cm}}c}
(\lambda x.t)u&=&t[u/x]&&(\beta)\\
\lambda x.tx&=& t&
\quad\text{if $x\notin t$}&(\eta)\\
\pi_1(\pair{u}{v})&=& u&&(\pi_1)\\
\pi_2(\pair{u}{v})&=& v&&(\pi_2)\\
\pair{\pi_1(u)}{\pi_2(u)}&=&u&&(\times)
\end{array}
$$

\paragraph{Type isomorphism:}

\bigskip

$$(t,u)\textrm{ s.t. }\begin{cases}
&\vdash t:A\fl B\\
&\vdash u: B\fl A\\
&\lambda x.t(u x)=\lambda x.u(t x)=\lambda x.x
\end{cases}
$$
}\vspace{-1cm}\\
\vspace{0.15cm}\\
\hline
\end{tabular}
\eec
\caption{Curry-style system F}
\end{figure}

\paragraph{Curry-style system F.} In the present work, we deal with type isomorphisms for
\textit{Curry-style} system F, \ie\ system F where the terms grammar
is simply the untyped $\lambda$-calculus' one. Although this system
appears to be less relevant than Church-style system F in proof-theory
(a term does not correspond exactly to one proof), it is actually more
accurate when we consider programming languages.  Indeed, in
Church-style system F, a term $t$ of type $\forall X. A$ will not have
the type $A[B/X]$: only $t\{B\}$ will be of this type; whereas in
Curry-style, a term $t$ of type $\forall X. A$ will have all the types
$A[B/X]$, which is more the idea induced by the notion of
polymorphism: the same function may be used with different types. The
typing rules and equalities of this language are presented on
figure~\ref{systF}.

\medskip


Compared with this system, Church-style system F has a different
grammar of terms:
$$t::=x\mid \lambda x^A.t\mid (tt)\mid\langle
t,t\rangle\mid\pi_1(t)\mid\pi_2(t)\mid\ \Lambda X.t\mid\ t\{A\}$$
different typing rules for the quantification:
\bec\AxiomC{$\Gamma\vdash t:A$}
\RightLabel{($\forall I$)}
\UnaryInfC{$\Gamma\vdash \Lambda X.t:\forall X.A$}
\DisplayProof if $X\notin\Gamma$
\quad
\AxiomC{$\Gamma\vdash t:\forall X.A$}
\RightLabel{($\forall E$)}
\UnaryInfC{$\Gamma\vdash t\{B\}:A[B/X]$}
\DisplayProof
\eec
and two additional equalities:
$$
\begin{array}{@{\hspace{4cm}}cccc@{\hspace{3.5cm}}c}
(\Lambda X.t)\{A\}&=&t[A/X]&&(\beta2)\\
\Lambda X.t\{X\}&=& t&\quad\text{if $X\notin t$}&(\eta2)\\
\end{array}
$$

As can be seen on the typing rules, a $\lambda$-term $t$ is of type $A$ if
there exists a term $\tilde{t}$ of Church-style system F such that $t$
is obtained from $\tilde{t}$ by erasing all the type indications (for example,
$\Lambda X.\lambda x^{\forall Y.Y}\lambda y^{Y}.x\{Y\}$ becomes $\lambda
x\lambda y.x$). In this case, we say that $t$ is the \textit{erasure}
of $\tilde{t}$.

The characterisation of type isomorphisms for Curry-style system F is
not directly reducible to the Church-style corresponding question: indeed,
types of the form $\forall X.A$ and $A$ with $X\notin A$ are not
equivalent in the Church-style setting, but they are in the
Curry-style one (where the isomorphism is realised by the identity). We prove in this paper that the distinction between
Church-style and Curry-style type isomorphisms can be resumed in one
new and non-trivial equation. To express it, one first have to recall
the definition of positive and negative type variables in a
formula\footnote{All along this article we will identify the notions
  of \textit{type} and \textit{formula} (according to the Curry-Howard
  correspondence).}:
\begin{Def}
If $A$ is a formula, its sets of \textbf{positive variables} $\posocc_A$ and
\textbf{negative variables} $\negocc_A$ are defined by:
\begin{itemize}
\item $\posocc_{X}=\{X\}$ , $\negocc_{X}=\emptyset$
\item $\posocc_{\bot}=\negocc_{\bot}=\emptyset$
\item $\posocc_{A\times B}=\posocc_{A}\cup\posocc_{B}$ , $\negocc_{A\times B}=\negocc_{A}\cup\negocc_{B}$
\item $\posocc_{A\fl B}=\negocc_{A}\cup\posocc_{B}$ , $\negocc_{A\fl B}=\posocc_{A}\cup\negocc_{B}$
\item $\posocc_{\forall X.A}=\posocc_{A}\ \backslash\ \{X\}$ ,
  $\negocc_{\forall X.A}=\negocc_{A}\ \backslash\ \{X\}$
\end{itemize}
We also define $FTV(A)=\posocc_A\cup\negocc_A$.
\end{Def}
The new equation is then the following:
$$\forall X.A\simeq_\varepsilon A[\forall Y.Y/X]\qquad\text{if }X\notin\negocc_A$$
It is true in Curry-style but false (in general) in Church-style
system F. Note that, although the isomorphism is realised by the
identity, the Church-style terms $t:\forall X.A\fl A[\forall Y.Y/X]$
and $u:A[\forall Y.Y/X]\fl\forall X.A$, from which we extract the
identity by erasing explicit types, are not trivial (they will
be explicitly described in the proof of theorem~\ref{charact} at the end of the paper). This is a difference
with Church-style system F, where type isomorphisms were exactly the
expected ones, even if proving that point was not an elementary task.

Type isomorphisms for Curry-style system F
are finally characterised by the following equational system:
\begin{align*}
A\times B&\eq B\times A&\\
A\times (B\times C)&\eq(A\times B)\times C&\\
A\fl (B\fl C)&\eq(A\times B)\fl C&\\
A\fl (B\times C)&\eq(A\fl B)\times(A\fl C)&\\
\forall X.\forall Y.A&\eq\forall Y.\forall X.A&\\
A\fl \forall X.B&\eq\forall X.(A\fl B)&\text{if $X\notin FTV(A)$}\\
\forall X.(A\times B)&\eq\forall X.A\times\forall X.B&\\
\forall X.A&\eq A[\forall Y.Y/X]&\text{if $X\notin\negocc_A$}
\end{align*}
The purpose of this paper is to prove correctness and completeness of
this characterisation by using a game model.

\paragraph{The model.} Models of second order calculi do not come about easily due to
impredicativity. Among the different possibilities, we choose
models based on game semantics because of their high degree of
adequation with the syntax: indeed, game semantics has been widely
used to construct fully complete models for various calculi, such
as PCF~\cite{ajm,ho}, $\mu$PCF~\cite{controlgames}, Idealized
Algol~\cite{iagames}, etc. This means that this semantics gives a
very faithful description of the behaviour of the syntax modulo
reduction rules in the system. And this is precisely what we need
to deal semantically with type isomorphisms: a model which is so
precise that it contains no more isomorphisms than the syntax.

The present paper introduces a game model for Curry-style system
F. This model was largely inspired by two preceding game semantics
works: the PhD thesis of Juliusz Chroboczek~\cite{phdjulius}, which
presents among others a game semantics for an untyped calculus that we
will almost copy-paste in this paper; and the game semantics model for
generic polymorphism by Samson Abramsky and Radha
Jagadeesan~\cite{gamesajf}, from which we will extract many ideas in
our context. Other game semantics models had an influence on our work:
Dominic Hughes gave the first game models of Church-style system
F~\cite{phdhughes} and introduced the notion of \textit{hyperforests}
that we reuse here; Andrzej Murawski and Luke Ong presented a simple
and efficient model for dealing with affine polymorphism~\cite{muraw},
and their presentation of moves inspired ours.

It shall be noticed that the design of our Curry-style game model is
actually very connected to the concepts present in the syntax: the
notion of erasure we introduce is of course reminiscent of
the erasure of types in a Church-like term to obtain a Curry-like
term. This is no surprise as we need a model describing very precisely
the syntax (that is why, in particular, one cannot be satisfied by an
interpretation of the quantification as an intersection or a greatest
lower bound). The specificities of (HON-)game semantics, as for
example the arborescent structure that interprets types, are however
decisive for our demonstration.


\section{General definitions}\label{general}

In this section we give general constructions that will apply on the
different grammars we use in the model. These constructions are
strongly related to usual HON-style games operations (cf.~\cite{ho}).


\subsection{Moves}

We consider the set of type variables $X,\ Y,\ \dots$ to be in
bijection with $\mathbb{N}\backslash\{0\}$, and we will further write
this set $\mc{X}=\{X_j\mid\ j>0\}$.

All along this article, we define several grammars of the form:
$$\mu::=\upa\mu\mid\ \dna\mu\mid\ \alpha_i\mu\mid\ j\qquad (i\in I,\ j\in \mathbb{N})$$
Let us note $\MM$ the set of words (often called \textbf{moves})
defined by this grammar.

Intuitively, the token $\upa$ (resp. $\dna$) corresponds to the right
side (resp. the left side) of an arrow type, the $\alpha_i$'s are
related to additional (covariant) connectors, the constants
$j\in\mathbb{N}\backslash\{0\}$ correspond to free type variables
$X_j$ and the constant $0$ corresponds either to bounded type
variables or to $\bot$.


\medskip

On such a grammar, we define automatically
a function $\lambda$ of \textbf{polarity}, with values in $\{\OO,\PP\}$:
\begin{itemize}
\item $\lambda(j)=\OO$
\item $\lambda(\upa\mu)=\lambda(\alpha_i\mu)=\lambda(\mu)$
\item $\lambda(\dna\mu)=\overline{\lambda}(\mu)$
\end{itemize}
where $\overline{\OO}=\PP$ and $\overline{\PP}=\OO$.

\medskip

We also introduce an \textbf{enabling relation} $\vdash\ \subseteq\MM\cup(\MM\times\MM)$:
\begin{itemize}
\item $\vdash j$
\item if $\vdash \mu$ then $\vdash \alpha_i\mu,$ and $\vdash \upa\mu$
\item if $\vdash\mu$ and $\vdash\mu'$ then $\upa\mu\vdash \dna\mu'$
\item if $\mu\vdash \mu'$ then $\alpha_i\mu\vdash \alpha_i\mu',$
  $\upa\mu\vdash \upa\mu'$ and $\dna\mu\vdash \dna\mu'$.
\end{itemize}
which induces a partial order $\leq$ for this grammar by reflexive
and transitive closure. If $\vdash\mu$ we say that $\mu$ is an
\textbf{initial move} (in which case $\lambda(\mu)=\OO$).

\subsection{Substitution}


As we want to deal with polymorphism,
we need some operations acting directly on the leafs $j$:
\begin{itemize}
\item a function $\sharp$ of \textbf{leaf extracting}:
\begin{itemize}
\item $\sharp(j)=j$
\item $\sharp(\upa\mu)=\sharp(\dna\mu)=\sharp(\alpha_i\mu)=\sharp(\mu)$
\end{itemize}
\item an operation of  \textbf{substitution} $\mu[\mu']$:
\begin{itemize}
\item $j[\mu']=\mu'$
\item $\upa\mu[\mu']=\upa(\mu[\mu'])$, $\dna\mu[\mu']=\dna(\mu[\mu'])$
  and $\alpha_i\mu[\mu']=\alpha_i(\mu[\mu'])$
\end{itemize}
\end{itemize}

We say that $\mu_1$ is a \textbf{prefix} of $\mu_2$ if there exists
$\mu'\in\mathcal{M}$ such that $\mu_2=\mu_1[\mu']$. This is denoted
$\mu_1\sqsubseteq^p\mu_2$.

\subsection{Plays and strategies}

\begin{Def}[justified sequence, play]
A \textbf{justified sequence} on a given grammar is a sequence
$s=\mu_1\dots \mu_n$ of moves, together with a partial function
$f:\{1,\dots,n\}\rightharpoonup\{1,\dots,n\}$ such that: if $f(i)$ is
not defined then $\vdash \mu_i$, and if $f(i)=j$ then $j<i$ and
$\mu_j\vdash \mu_i$: in this case we say that $\mu_j$ \textbf{justifies}
$\mu_i$.

A \textbf{play} on a grammar is a justified sequence $s=\mu_1\dots
\mu_n$ on this grammar such that: for every $1\leq i\leq n-1$, if $\lambda(\mu_i)=\PP$ then
$\lambda(\mu_{i+1})=\OO$ and if $\lambda(\mu_i)=\OO$ then
$\lambda(\mu_{i+1})=\PP$ \textit{and}
$\sharp(\mu_i)=\sharp(\mu_{i+1})$.

We note $\mathbb{E}$ the set of plays of even length.
If $s$ and $t$ are two plays, we note $t\preceq s$ if
$t$ is a prefix of $s$.
\end{Def}

The definition of a play implies that if $s\mu\nu$ is an even-length play
then $\sharp(\mu)=\sharp(\nu)$. This will be a very significant property in our
model.

\begin{Def}[strategy]
A \textbf{strategy} $\sigma$ on a given grammar is a non-empty set of
even-length plays, which is closed under even-length prefix and
deterministic: if $s\mu$ and $s\nu$ are two plays of $\sigma$ then
$s\mu=s\nu$.
\end{Def}

\begin{Def}[view, innocence]
Let $s$ be a play on a grammar, we define its \textbf{view} $\view{s}$
by:
\begin{itemize}
\item $\view{\varepsilon}=\varepsilon$
\item $\view{s\mu}=\view{s}\mu$ if $\lambda(\mu)=\PP$
\item $\view{s\mu}=\mu$ if $\vdash \mu$
\item $\view{s\mu t\nu}=\view{s}\mu\nu$ if $\lambda(\nu)=\OO$ and
  $\mu$ justifies $\nu$
\end{itemize}

A strategy $\sigma$ is called \textbf{innocent} if, for every play
$s\nu$ of $\sigma$, the justifier of $\nu$ is in $\ulcorner
s\urcorner$, and if we have: if $s\mu\nu\in\sigma$, $t\in\sigma$,
$t\mu$ is a play and $\ulcorner s\mu\urcorner=\ulcorner t\mu\urcorner$
then $t\mu\nu\in\sigma$.
\end{Def}

\begin{Def}[bi-view]
A \textbf{bi-view} on a given grammar is a justified sequence
$s=\mu_1\dots \mu_n$ (with $n\geq 1$) such that any move is justified by its
predecessor. The set of bi-views is denoted $\mc{BV}$.
\end{Def}

\subsection{Composition}

Composition is usually defined between arenas of the form $A\fl B$ and
$B\fl C$. We are going to define it in a context where arenas do not
explicitly exist, but are however represented by the tokens $\upa$
and $\dna$.

\begin{Def}[shape]
Let $\zeta\in(\{\upa,\dna\}\cup\{\alpha_i\}_{i\in I})^*$, a move $\mu$
is said to be of \textbf{shape} $\zeta$ if 
$\zeta 0\sqsubseteq^p\mu$.

Let $\Sigma$ be a finite set of elements $\zeta_j\in(\{\upa,\dna\}\cup\{\alpha_i\}_{i\in I})^*$.
A justified sequence is said to be of shape $\Sigma$ if each of its moves is of shape
$\zeta_j$ for some $j$. A strategy is of shape $\Sigma$ if each of its plays is of shape $\Sigma$.

In the case where $\Sigma=\{\upa,\dna\}$, we say that the justified sequence
(or the strategy) is of \textbf{arrow shape}.
\end{Def}

Consider a justified sequence $s=\mu_1\dots \mu_n$, we define the
sequence $s\restr_\zeta$ as the restriction of $s$ to the moves of
shape $\zeta$ where the prefix $\zeta$ has been erased, and the
pointers are given as follows: if $\mu_i=\zeta\mu'_i$ is justified by
$\mu_j=\zeta\mu'_j$ in $s$, then the corresponding occurrence of $\mu'_i$
is justified by $\mu'_j$

Consider $\zeta,\xi\in(\{\upa,\dna\}\cup\{\alpha_i\}_{i\in I})^*$ such
that neither of the two is a prefix of the other. Let us define the
sequence $s\restr_{\zeta,\xi}$: first we consider $s'$, the
restriction of $s$ to the moves of shape $\zeta$ and the moves of
shape $\xi$ hereditarily justified by a move of shape $\zeta$.
$s\restr_{\zeta,\xi}$ is the sequence $s'$ where the prefix
$\zeta$ has been replaced by $\upa$ where it appears, the prefix
$\xi$ has been replaced by $\dna$ where it appears, and the pointers
are given as follows: if $\mu_i=\zeta\mu'_i$ (resp. $\mu_i=\xi\mu'_i$)
is justified by $\mu_j=\zeta\mu'_j$ (resp. $\mu_j=\xi\mu'_j$) in $s$,
then the corresponding occurrence of $\upa\mu'_i$ (resp. $\dna\mu'_i$)
is justified by $\upa\mu'_j$ (resp. $\dna\mu'_j$); and if
$\mu_i=\xi\mu'_i$ is hereditarily justified by a move
$\mu_j=\zeta\mu'_j$ in $s$, then the corresponding occurrence of
$\dna\mu'_i$ is justified by the corresponding occurrence of
$\upa\mu'_j$ iff $\vdash \mu'_i$ and $\vdash \mu'_j$.

\begin{Def}[interacting sequence, composition]
An \textbf{interacting sequence} $s=\mu_1\dots \mu_n$ is a justified sequence of shape
$\{\upa,\dna\upa,\dna\dna\}$ such that $s\restr_{\upa,\dna\upa}$,
$s\restr_{\dna\upa,\dna\dna}$ and $s\restr_{\upa,\dna\dna}$ are plays.
The set of interacting sequences is denoted \textbf{Int}.

Suppose we have two strategies 
$\sigma$ and $\tau$. We call \textbf{composition} of $\sigma$
and $\tau$ the set of plays
$$\sigma;\tau=\{u\restr_{\upa,\dna\dna} \mid u\in \textbf{Int},\
u\restr_{\upa,\dna\upa}\in\tau\ \textit{and}\
u\restr_{\dna\upa,\dna\dna}\in\sigma\}$$
\end{Def}

$\sigma;\tau$ is a strategy: this can be proven like in the standard
HON game model. Moreover if $\sigma$ and $\tau$ are innocent then
$\sigma;\tau$ is innocent.

\begin{Def}[totality on a shape]
Let $\sigma$ be a strategy and
$\zeta\in(\{\upa,\dna\}\cup\{\alpha_i\}_{i\in I})^*$. We say that
$\sigma$ is \textbf{total} on the shape $\zeta$ if, for every play
$s\in\sigma$ of shape $\zeta$, for every move $\mu$ such that $s\mu$
is a play of shape $\zeta$, there exists a move $\nu$ of shape $\zeta$
such that $s\mu\nu\in\sigma$.
\end{Def}\label{fgen}

\subsection{Presentation of the Curry-style model}


Our model is defined through three grammars:
\begin{itemize}
\item $\mathbb{X}$ is the grammar of \textbf{untyped moves} which
  generate the untyped model to interpret untyped lambda-terms
\item $\mathbb{A}$ is the grammar of \textbf{occurrences} which are
  used for the interpretation of formulas
\item $\mathbb{M}$ is the grammar of \textbf{typed moves} which
  generate an interpretation of the terms of Church-style system F.
\end{itemize}

\medskip

The interpretation of Curry-style system F in the model will be as follows: 
\begin{itemize}
\item a type $A$ will be interpreted as a \textit{game} (also denoted $A$), \ie\ a specific
  structure based on the grammar $\mathbb{A}$
\item a term $t$ of type $A$ will be interpreted as a strategy $\sigma$ on the grammar
  $\mathbb{X}$, with the condition that this strategy is the
  \textbf{erasure} of a strategy $\tilsig$, defined on the grammar
  $\mathbb{M}$ and played on the game $A$ (this will be denoted $\tilsig::A$)
\item two additional properties are required:
  \textbf{hyperuniformity} which applies on $\sigma$, and
  \textbf{uniformity} which applies on $\tilsig$.
\end{itemize}

\medskip

In what follows, we first define the untyped model to interpret
untyped lambda-terms, then we define games and typed strategies on
games, and finally we introduce the notion of erasure and prove that
we have a model of Curry-style system F. Next we prove, using this
model, our result on type isomorphisms.

\section{The untyped model}\label{dunt}

In this section we give a semantics for the untyped $\lambda$-calculus
with binary products, \ie\ for the calculus of
figure~\ref{systF} restricted to the language of terms with their
reduction rules.

The untyped model that we present below has been
defined by Julius Chroboczek in his PhD thesis~\cite{phdjulius}. Our
definition is formally a little bit different from Chroboczek's one,
but the substance of the work is the same.

\subsection{Untyped moves}

The grammar of \textbf{untyped moves} is the following:
$$x::=\upa x\mid\ \dna x\mid\ r x\mid\ l x\mid\ j\qquad(j\in\mathbb{N})$$
The set of untyped moves is denoted $\mathbb{X}$.

The justified sequences, plays and strategies induced by this grammar
will be called \textit{untyped} justified sequences, plays and
strategies.

\subsection{Basic strategies}


We define the following strategies:
\begin{itemize}
\item \textbf{identity}:
$$\id=\{s\in\mathbb{E}\mid\ s\text{ of arrow shape }\text{ and }\forall
  t\in\mathbb{E},t\preceq s\Rightarrow t\restr_\upa=t\restr_\dna\}$$
\item \textbf{projections}:
$$\pi_r=\{s\in\mathbb{E}\mid\ s\text{ of shape }\{\upa,\dna r,\dna
  l\}\text{ and }\forall t\in\mathbb{E},t\preceq s\Rightarrow
  t\restr_\upa=t\restr_{\dna r}\}$$
$$\pi_l=\{s\in\mathbb{E}\mid\ s\text{ of shape }\{\upa,\dna r,\dna
  l\}\text{ and }\forall t\in\mathbb{E},t\preceq s\Rightarrow
  t\restr_\upa=t\restr_{\dna l}\}$$
\item \textbf{evaluation}:
$$\textit{eval}=\{s\in\mathbb{E}\mid\ s\text{ of shape }\{\upa,\dna l\upa,\dna l\dna,\dna
r\}\text{ and }\forall t\in\mathbb{E},t\preceq s\Rightarrow t\restr_\upa=t\restr_{\dna
l\upa}\wedge t\restr_{\dna r}=t\restr_{\dna l\dna}\}$$
\end{itemize}


We also define three basic operations on strategies:
\begin{itemize}
\item \textbf{pairing without context}: if $\sigma$ and $\tau$ are two
  strategies,
$$\pair{\sigma}{\tau}_a=\{s\in\mathbb{E}\mid\ s\text{ of shape }\{r,l\}\text{ and }
  s\restr_{l}\ 
  \in\sigma\text{ and } s\restr_{r}\ \in\tau\}$$
\item \textbf{pairing with context}: if $\sigma$ and $\tau$ are two strategies of
  arrow shape,
$$\pair{\sigma}{\tau}_b=\{s\in\mathbb{E}\mid\ s\text{ of shape }\{\upa
  r,\upa l,\dna\}\text{ and }
  s\restr_{\upa l,\dna}\
  \in\sigma\text{ and } s\restr_{\upa r,\dna}\ \in\tau\}$$
\item \textbf{abstraction}: if $\sigma$ is a strategy of shape $\{\upa,\dna r,\dna l\}$,
$\Lambda(\sigma)$ is the strategy of shape
$\{\upa\upa,\upa\dna,\dna\}$ which is deduced from $\sigma$ by
replacing each move $\upa x$ by $\upa\upa x$, each move
$\dna rx$ by $\upa\dna x$ and each move $\dna lx$ by
$\dna x$.
\end{itemize}


\subsection{Hyperuniformity}

We have enough material to define an untyped model. However, our use
of untyped strategies in the Curry-style model forces us to impose new
requirements: for example, consider the formula $X_1\fl X_1$. It would
be reasonable to think that the innocent strategy $\sigma$ whose set
of views is $\{\varepsilon,\upa1\cdot\dna1\}$ has this type. However,
because we deal with a Curry-style model, any strategy of type $X_1\fl
X_1$ should also have the type $\forall X_1.X_1\fl X_1$, and thus
$A\fl A$ for any $A$, and should be able to do a copycat between
the left and the right side of the arrow.

This is the meaning of the notion of \textbf{hyperuniformity} defined
below.

\begin{Def}[copycat extension of an untyped play]
Let $s=x_1\dots x_n$ be an untyped play, $x_i$ an $\OO$-move of $s$
and $v=y_1\dots y_p\in\mc{BV}$. Suppose $s=s_1x_{i}x_{i+1}s_2$.  The
\textbf{copycat extension} of $s$ at position $i$ with parameter
$v$ is the untyped play $s'=\textit{cc}^s(i,v)$, defined by :
\begin{itemize}
\item $s'=s_1x_{i}[y_1]x_{i+1}[y_1]s_2$ if $p=1$
\item $s'=s_1x_{i}[y_1]x_{i+1}[y_1]x_{{i}+1}[y_2]x_{i}[y_2]\dots
x_{{i}+1}[y_p]x_{i}[y_p]$ if $p$ even
\item $s'=s_1x_{i}[y_1]x_{i+1}[y_1]x_{{i}+1}[y_2]x_{i}[y_2]\dots
x_{i}[y_p]x_{i+1}[y_p]$ if $p>1$ and $p$ odd
\end{itemize}
\end{Def}




\begin{Def}[hyperuniform strategy]
An untyped strategy $\sigma$ is called \textbf{hyperuniform} if it is
innocent and if, for any play $s\in\sigma$, any copycat extension of
$s$ is in $\sigma$.
\end{Def}

\begin{Lemme}\label{hyperunif1}
The identity strategy, the projections and the evaluation strategy are
hyperuniform. If $\sigma$ and $\tau$ are hyperuniform then
$\pair{\sigma}{\tau}$ and $\Lambda(\sigma)$ are hyperuniform.
\end{Lemme}

The preceding lemma is straightforward. The interesting case is
composition:

\begin{Lemme}\label{hyperunif2}
If $\sigma$ and $\tau$ are hyperuniform then $\sigma;\tau$ is hyperuniform.
\end{Lemme}

\proof Let us consider a play $s=x_1\dots
x_p\in\sigma;\tau$, an $\OO$-move $x_i$ of $s$ and a
bi-view $v=y_1\dots y_q$.  We have to prove that
$s'=\textit{cc}^s(i,v)$ belongs to $\sigma;\tau$.

There exists a justified sequence $u$ such that
$u\restr_{\upa,\dna\dna}=s$, $u\restr_{\dna\upa,\dna\dna}\in\sigma$
and $u\restr_{\upa,\dna\upa}\in\tau$. If $u=t_1x_{i}b_1\dots
b_qx_{i+1}t_2$, we build a new justified sequence $U$ depending on the
value of $p$ :
\begin{itemize}
\item if $p=1$, $U=t_1x_{i}[y_1]b_1[y_1]\dots b_q[y_1]x_{i+1}[y_1]
t_2$
\item if $p$ even,\\
$U=t_1x_{i}[y_1]b_1[y_1]\dots b_q[y_1]x_{i+1}[y_1]
x_{i+1}[y_2]b_q[y_2]\dots b_1[y_2]x_{i}[y_2]\dots\dots
x_{i+1}[y_p]b_q[y_p]\dots b_1[y_p]x_{i}[y_p]$
\item if $p$ odd and $p>1$,\\
$U=t_1x_{i}[y_1]b_1[y_1]\dots b_q[y_1]x_{i+1}[y_1]
x_{i+1}[y_2]b_q[y_2]\dots b_1[y_2]x_{i}[y_2]\dots\dots
x_{i}[y_p]b_1[y_p]\dots b_q[y_p]x_{i+1}[y_p]$
\end{itemize}

We have $U\restr_{\dna\upa,\dna\dna}\in\sigma$ and
$U\restr_{\upa,\dna\upa}\in\tau$ by hyperuniformity of $\sigma$ and
$\tau$. So, $U\restr_{\upa,\dna\dna}=s'\in\sigma;\tau$.

\cqfd


\subsection{Semantics of the untyped $\lambda$-calculus with binary products}


We now present the interpretation of the untyped calculus.
Instead of directly interpreting terms, we interpret sequents of the
form $\Gamma\vdash t$, where $t$ is a term and $\Gamma$ is simply a
list of variables that includes the free variables occurring in $t$.

The interpretation is as follows:
\begin{gather*}
\trad{x\vdash x}=\id\\
\phantom{ceeeeeeeentre}\trad{\Gamma,x\vdash x}=\pi_r\qquad\text{if $\Gamma\neq \emptyset$}\\
\trad{\Gamma,y\vdash x}=\pi_l;\trad{\Gamma\vdash x}\\
\trad{\Gamma\vdash \lambda x.t}=\Lambda(\trad{\Gamma,x\vdash t})\\
\trad{\Gamma\vdash (t u)}=\langle\trad{\Gamma\vdash t},\trad{\Gamma\vdash u}\rangle_{a(\Gamma)};\textit{eval}\\
\trad{\Gamma\vdash \langle t,u\rangle}=\langle\trad{\Gamma\vdash t},\trad{\Gamma\vdash u}\rangle_{a(\Gamma)}\\
\trad{\Gamma\vdash \pi_1(t)}=\trad{\Gamma\vdash t};\pi_l\\
\trad{\Gamma\vdash \pi_2(t)}=\trad{\Gamma\vdash t};\pi_r
\end{gather*}
with $a(\Gamma)=a$ if $\Gamma=\emptyset$ and $a(\Gamma)=b$ otherwise.

\bigskip

From lemmas~\ref{hyperunif1} and~\ref{hyperunif2} we derive:

\begin{Lemme}
Let $t$ be a term whose free variables are contained in the list $\Gamma$, then 
$\trad{\Gamma\vdash t}$ is a hyperuniform strategy.
\end{Lemme}

\begin{Prop}\label{jul}
If two terms $t$ and $u$ are equal up to the equalities of the language, and if all their free variables 
are contained in the list $\Gamma$,
then $\trad{\Gamma\vdash t}=\trad{\Gamma\vdash u}$.
\end{Prop}

See~\cite{phdjulius} for the proof of the equivalent proposition in Chroboczek's
setting.\label{funt}

\section{Games}\label{dgames}

\subsection{Interpretation of a formula}

In this section we introduce the notion of \textbf{game}\footnote{The
  denomination \textit{arena} would also fit, but we wanted to stress
  the fact that our games are not trees like HON-arenas, but just
  partial orders.}, the structure that will interpret Curry-style
types. This structure is very similar to the one presented
in~\cite{gamesajf}.

\medskip

We define the following grammar of \textbf{occurrences}:
$$a::=\upa a\mid\ \dna a\mid\ r a\mid\ l a\mid\ \star a\mid\ j\qquad
(j\in\mathbb{N})$$
The set of all occurrences is denoted $\mathbb{A}$.

We define a translation $\mc{E}$ from $\mathbb{A}$ to $\mathbb{X}$:
$\mc{E}(a)$ is obtained by erasing all the tokens $\star$ in
$a$. Inductively:
\begin{itemize}
\item $\mc{E}(i)=i$
\item $\mc{E}(\star a)=\mc{E}(a)$
\item $\mc{E}(\alpha a)=\alpha\mc{E}(a)$ if $\alpha\in\{\upa,\dna,r,l\}$.
\end{itemize}

\bigskip


The \textbf{syntactic tree} of a formula $A$ is a tree with nodes
labelled by type connectors ($\fl,\times,\forall$) or integers, edges
labelled by the tokens $\upa,\dna,r,l,\star$, and possibly some arrows
linking a leaf to a node. It is defined as follows:
\begin{itemize}
\item $T_{\bot}$ is reduced to a leaf $0$
\item $T_{X_i}$ is reduced to a leaf $i$
\item $T_{A\fl B}$ consists in a root $\fl$ with the two trees $T_A$
  and $T_B$ as sons; the edge between $\fl$ and $T_A$ (resp. $T_B$) is
  labelled $\dna$ (resp. $\upa$)
\item $T_{A\times B}$ consists in a root $\times$ with the two trees
  $T_A$ and $T_B$ as sons; the edge between $\times$ and $T_A$
  (resp. $T_B$) is labelled $l$ (resp. $r$)
\item $T_{\forall X_i.A}$ consists in a root $\forall$ with the tree
  $T$ as unique son, where $T$ is deduced from $T_A$ by linking each
  of its leafs labelled by $i$ to its root, and relabelling these
  leafs by $0$; the edge between $\forall$ and $T$ is labelled
  $\star$.
\end{itemize}

A maximal branch in a syntactic tree is a path from the root to a
leaf; it will be described by the sequence of labels of its edges,
with the index of the leaf at the end of the sequence. Such a maximal
branch is then an occurrence.

The set $\occ_A$ of occurrences of a formula $A$ is the set of maximal
branches of $T_A$. We define a function of \textbf{linkage}
$\LLL_A:\occ_A\fl\mathbb{A}\cup\{\dag\}$ as follows: if the leaf
reached by the maximal branch $a$ is linked to a node $c$, then
$\LLL_A(a)$ is the sequence of labels of the edges we cross to reach
$c$ starting from the root, with a $0$ at the end; otherwise,
$\LLL_A(a)=\dag$.

The structure $(\occ_A,\LLL_A)$ will be called a \textbf{game}. It
will also be denoted $A$, with no risk of confusion.

\bigskip

\paragraph{Example:}
The type $A=\forall X_1.(X_1\fl((\forall X_2.X_2)\fl
(X_3\times\bot)))$ has as set of occurrences:
$$\occ_A=\{\star\dna0\ ,\ \star\upa\dna\star0\ ,\ \star\upa\upa l3\ ,\ \star\upa\upa
r0\}$$
And its function of linkage is given by:
$$\begin{cases}
\begin{array}{lcc}
\mc{L}_A(\star\dna0)&=&\star0\\
\mc{L}_A(\star\upa\dna\star0)&=&\star\upa\dna\star0\\
\mc{L}_A(\star\upa\upa l3)&=&\dag\\
\mc{L}_A(\star\upa\upa r0)&=&\dag
\end{array}&
\end{cases}$$

\bigskip


\begin{Def}[game]\label{defgame}
A \textbf{game} $A$ is defined by a finite non-empty set $\occ_A\subseteq\mathbb{A}$
and a function of \textbf{linkage}
$\LLL_A:\occ_A\fl\mathbb{A}\cup\{\dag\}$ satisfying the following
conditions:
\begin{itemize}
\item $\occ_A$ is \textbf{coherent}: for every
  $a\in\occ_A$, either $\vdash a$ or $\exists a'\in\occ_A,\ a'\vdash a$
\item $\occ_A$ is \textbf{non-ambiguous}: $\forall a,a'\in\occ_A$, if
  $\mc{E}(a)\sqsubseteq^p \mc{E}(a')$ then $a=a'$
\item for every $a\in\occ_A$, either $\LLL_A(a)=\dag$ or
  $\LLL_A(a)=a'[\star 0]\sqsubseteq^p a$ for some $a'\in\mathbb{A}$
\item for every $a\in\occ_A$, if $\sharp(a)\neq 0$ then $\LLL_A(a)=\dag$
\end{itemize}
The set of games is denoted $\mc{G}$.
\end{Def}


We stress the fact that the set $\occ_A$ shall not be empty: this will
be a crucial point in our proofs.

\begin{Def}[auxiliary polarity]
Given a game $A$, we define its \textbf{auxiliary polarity} as a
partial function $\paux_A:\occ_A\rightharpoonup\{\OO,\PP\}$ by:
$\paux_A(c)=\lambda(\LLL_A(c))$ if $\LLL_A(c)\neq \dag$, otherwise it
is undefined.
\end{Def}


\subsection{Alternative, inductive interpretation of a formula}

We define the following constructions on games:
\begin{minilist}
\item[\textbf{(atoms)}] $\bot=(\{0\},0\mapsto\dag)$ \qquad $X_i=(\{i\},i\mapsto\dag)$ for $i>0$.
\item[\textbf{(product)}] if $A,B\in\mc{G}$, we define $A\times B$ by:
\begin{itemize}
\item
  $\occ_{A\times B}=\{l a\mid\ a\in\occ_A\}\cup\{r
  b\mid\ b\in\occ_B\}$
\item $\LLL_{A\times B}(l a)=\begin{cases}
  \dag&\text{if }\LLL_A(a)=\dag\\ l\LLL_A(a)&\text{otherwise}
\end{cases}$\qquad$\LLL_{A\times B}(r b)=\begin{cases}
\dag&\text{if }\LLL_B(b)=\dag\\ r\LLL_B(b)&\text{otherwise}
\end{cases}$
\end{itemize}
\item[\textbf{(arrow)}] if $A,B\in\mc{G}$, we define $A\fl B$ by:
\begin{itemize}
\item
  $\occ_{A\fl B}=\{\dna a\mid\ a\in\occ_A\}\cup\{\upa
  b\mid\ b\in\occ_B\}$
\item $\LLL_{A\fl B}(\dna a)=\begin{cases}
  \dag&\text{if }\LLL_A(a)=\dag\\ \dna\LLL_A(a)&\text{otherwise}
\end{cases}$\qquad$\LLL_{A\fl B}(\upa b)=\begin{cases}
\dag&\text{if }\LLL_B(b)=\dag\\ \upa\LLL_B(b)&\text{otherwise}
\end{cases}$
\end{itemize}
\item[\textbf{(quantification)}] if $A\in\mc{G}$ and $i>0$, we define
  $\forall X_i.A$ by:
\begin{itemize}
\item
  $\occ_{\forall X_i.A}=\{\star a\mid\ a\in \occ_A\wedge \sharp(a)\neq
  i\}\cup\{\star a[0]\mid\ a\in \occ_A\wedge \sharp(a)= i\}$ 
\item $\LLL_{\forall X_i.A}(\star a)=\begin{cases}
\dag&\text{if }\LLL_A(a)=\dag\\
\star\LLL_A(a)&\text{otherwise}
\end{cases}$
\qquad
$\LLL_{\forall X_i.A}(\star a[0])=\star 0$
\end{itemize}
\end{minilist}

This gives rise to an inductive interpretation of a formula, which
coincides with the one defined from the syntactic tree.

\bigskip

Finally, we define an operation of substitution on games:

\begin{Def}[substitution]
Let $A,B\in\mc{G}$. The \textbf{substitution} of $X_i$ by $B$ in $A$
is the game $A[B/X_i]$ defined by:
\begin{itemize}
\item $\occ_{A[B/X]}=\{a\in\occ_A\mid\ \sharp(a)\neq
  i\}\cup\{a[b]\mid\ a\in\occ_A\wedge \sharp(a)=i\wedge b\in\occ_B\}$
\item $\LLL_{A[B/X]}(a)=\LLL_A(a)$ and $\LLL_{A[B/X]}(a[b])=\begin{cases}
\dag&\text{if }\LLL_B(b)=\dag\\
a[\LLL_B(b)]&\text{otherwise}
\end{cases}$
\end{itemize}
\end{Def}

One can check that this coincides with the operation of substitution
on formulas.\label{fgames}

\section{The typed model}

\subsection{Moves and strategies on a game}\label{dtyped}

We are now going to describe how we can play in a game. We will take
advantage of the way we have defined games: whereas in many second
order game models like~\cite{phdhughes} or \cite{lataillade} moves have
a complex structure, here they will be easy to derive from
$\occ_A$ and $\mc{L}_A$.

As in~\cite{gamesajf}, the intuition is
that a move in $A$ can either be built directly from an occurrence of
$\occ_A$, or it can be decomposed as $m_1[m_2]$, where $m_1$ is built
from an occurrence of $\occ_A$ and $m_2$ is a move in another game $B$
which substitutes a quantifier.

Note that the moves and strategies defined this way do not constitute
the morphisms of our model, but they will be used as interpretations
of Church-style terms.

\bigskip

We introduce the grammar of \textbf{typed moves}:
$$m::=\upa m\mid\ \dna m\mid\ r m\mid\ l m\mid\ \star^B m\mid\ j\qquad (B\in\mc{G},j\in\mathbb{N})$$
These moves form the set $\mathbb{M}$.

The operation of \textbf{anonymity} $\mc{A}:\mathbb{M}\fl\mathbb{A}$
erases the game indication in a typed move:
\begin{itemize}
\item $\AAA(i)=i$ for $i\geq0$
\item $\AAA(\star^Am)=\star\AAA (m)$
\item $\AAA(\alpha m)=\alpha\AAA (m)$ for $\alpha\in\{r,l,\upa,\dna\}$.
\end{itemize}

For $m\in\mathbb{M}$ and $a\in\mathbb{A}$, we define a partial operation of
\textbf{formula extraction} $\frac{m}{a}$
by:
\begin{itemize}
\item $\frac{\star^Bm}{\star0}=B$
\item if $\frac{m}{a}$ is defined, $\frac{\star^Bm}{\star
  a}=\frac{\alpha m}{\alpha a}=\frac{m}{a}$ where 
$\alpha\in\{\upa,\dna,r,l\}$
\end{itemize}

\begin{Def}[moves of a game]
Let $A$ be a game. Its set of \textbf{moves} $\MM_A\subseteq
\mathbb{M}$ is given by defining the relation $m\in\MM_A$ by
induction on $m$:
\begin{itemize}
\item if $\AAA(m)=a\in\occ_A$ and $\LLL_A(a)=\dag$ then $m\in\MM_A$
\item if $m=m_1[m_2]$, where $\AAA(m_1)=a\in\occ_A$, $\LLL_A(a)\neq\dag$ and
$m_2\in\MM_B$ with $B=\frac{m_1}{\LLL_A(a)}$, then $m\in\MM_A$.
\end{itemize}
\end{Def}

This definition is well-defined, because in the second case we
necessarily have at least one token $\star^B$ in $m_1$, so the size of
$m_2$ is strictly smaller than the size of $m_1[m_2]$: that is why we
say that the definition is inductive.

\medskip

\paragraph{Example:} Let us recall the type $A=\forall
X_1.(X_1\fl((\forall X_2.X_2)\fl (X_3\times\bot)))$ of the preceding
example. One possible way to ``play a move'' in this
game\footnote{This notion is related to the idea of \textbf{evolving
game} introduced in~\cite{muraw} and reused in~\cite{lataillade}.} is
to instantiate the variable $X_1$ with a type $B$ (take $B=\bot\times
X_3$ for example), then to go on the left side of the first arrow and
to play a move of $B$.

This corresponds to a move like $m=\star^{B}\dna r3$. One can check
with the definition that this move indeed belongs to $\MM_A$:
$m=m_1[m_2]$ with $m_1=\star^{B}\dna0$ and
$m_2=r3$. $\AAA(m_1)=\star\dna0\in\occ_A$,
$\LLL_A(\star\dna0)=\star0$ and
$\frac{\star^{B}\dna0}{\star0}=B$. Moreover,
$\AAA(m_2)=r3\in\occ_B$ and $\LLL_B(m_2)=\dag$ so $m_2\in\MM_B$
(first case of the definition). So, $m\in\MM_B$ (second case of the
definition).

Intuitively, we have the following: 
\begin{itemize}
\item $m_1$ is the part of the move played in $A$, and $c=\AAA(m_1)$
  is the corresponding occurrence
\item $\LLL_a(c)$ indicates where the interesting quantifier has been
instantiated
\item $\frac{m_1}{\LLL_A(c)}=B$ indicates by which game it has been
instantiated
\item $m_2$ is the part of the move played in $B$.
\end{itemize}

\smallskip

\begin{Def}[justified sequence, play on a game]
Let $A$ be a game and $s$ be a play (resp. a justified sequence) on
the grammar $\mathbb{M}$. If every move of $s$ belongs to $\MM_A$,
then we say that $s$ is a play (resp. a justified sequence) on the
game $A$.  The set of plays on the game $A$ is denoted $\mc{P}_A$.
\end{Def}

\smallskip

\paragraph{Example:} Let us consider the play $s=\star^B\upa\upa
l3\cdot\star^B\dna r3$ with $B=\bot\times X_3$. This is of course a
play \textit{in} $A=\forall X_1.(X_1\fl(\forall X_2.X_2)\fl
(X_3\times\bot))$.

What is interesting to notice is that, if for
example $C=X_3\times\bot$, then the sequence $s'=\star^C\upa\upa
l3\cdot\star^B\dna r3$ is not a play because it is not a
justified sequence: indeed, one must have $B=C$ if we want
$m_2=\star^B\dna r3$ to be justified by $m_1=\star^C\upa\upa
l3$.

More generally, for any move $m$ in a play $s$ which contains the
token $\star^B$, there is a sequence of moves $m_1,\dots,m_n$ that
also contains the token $\star^B$ at the same place, with $m_n=m$ and
$m_i$ justifies $m_{i+1}$ for $1\leq i<n$. If this sequence is chosen
to be of maximal length, then $m_1$ is the minimal hereditarily
justifier of $m$ which contains the token $\star^B$: it is the first
time that it appears (at the right place). We will say that $B$ is
played by $\lambda(m_1)$ at the \textbf{level} of $m_1$. Note that
$\lambda(m_1)=\paux_A(m)$.

\bigskip

One can formalise this definition:

\begin{Def}[level]
If a move $m$ in a play $s\in\mc{P}_A$ contains the token $\star^B$,
then it can be written $m=m_0\star^B[m_1]$. We say that $B$ is played
(by $\lambda(m_0)$) at the \textbf{level} of $m$ if $m_1$ does not
contain the token $\dna$.
\end{Def}


\smallskip

Typed strategies are defined as expected:

\begin{Def}[strategy on a game]
Let $\sigma$ be a strategy on the grammar $\mathbb{M}$, we say that
$\sigma$ is a strategy on $A$ and we note $\sigma::A$ if any play of
$\sigma$ belongs to $\mc{P}_A$. We say that $\sigma$ is a typed
strategy in this case.
\end{Def}

Strategies on games have to be understood as
interpretations\footnote{We chose not to explicit this interpretation
because we do not need it; one could also prove that we have a model
of Church-style system F, but it is not an important question here.}
of Church-style system F terms; they will be used in the Curry-style
model because we have to express in the model the fact that a
well-typed Curry-style term is the erasure of a well-typed
Church-style term.

\subsection{Uniformity}


In~\cite{lataillade}, we saw that strategies defined as generally as
possible were not able to capture exactly the type isomorphisms of the
syntax, because they were generating too many isomorphisms in the
model. That is why we introduced a notion of \textit{uniformity},
which restrained the behaviour of strategies (in order to avoid
confusion, we will call \textit{weak uniformity} the notion of
uniformity defined in~\cite{lataillade}; by the way, weak uniformity
plays no role in the present model).

The situation is similar here: we are not able to derive the
characterisation of Curry-style type isomorphisms if the well-typed
Church-style terms are interpreted by the (typed) strategies defined
above. So we introduce a notion of \textbf{uniformity} on these
strategies.

The intuition of this notion is the following: consider an
$\eta$-long, $\beta$-normal term $t$ of the Church-style system F, and
suppose $\vdash t:\forall X.A$. The term $t$ has the form $t=\Lambda
X.t'$ with $\vdash t':A$: so it behaves like if it was instantiating the
quantifier ($\forall X$) with a variable ($X$). More generally, the
terms of the Church-style system F should be interpreted by strategies where, each
time $\OO$ has to play a game, he gives a variable game $X_i$.

But these strategies (that we will call \textbf{symbolic}) do not
compose: in the Church-style syntax, this corresponds to the fact that
the term $\vdash t:\forall X.A$ can be instantiated at any type $B$
through the operation $t\mapsto t\{B\}$, and so the term $t$ can be
extended to any type $A[B/X]$. In the model, this means that the
symbolic strategy interpreting $t$ must be extensible to a more
complete strategy, where $\OO$ can play any game he wants.  This
extension consists in playing copycat plays between the different
occurrences of the variables $X$ (like in the syntax, the $\eta$-long
$\beta$-normal form of $t\{B\}$ is generated from $t$ through
$\eta$-expansions), that is why it is called the \textbf{copycat
extension}.

To sum up, a uniform strategy will be a symbolic strategy extended by
copycat extension. This idea has to be related with the strategies of
Dominic Hughes~\cite{phdhughes} and, above all, with Murawski's notion of
\textit{good strategies}~\cite{muraw}. The notion of weak uniformity discussed
above is an analogous, but less restrictive, condition: uniformity
implies weak uniformity. Finally, uniformity has of course a strong connection
with hyperuniformity: the two notions express analogous ideas, but hyperuniformity
applies on untyped strategies, whereas uniformity is formulated in a
typed context, and then requires more cautiousness.

\bigskip

In the following definition, $\mc{BV}(A)$ stands for the set of bi-views
in a game $A$, and $m[B/j]$ (resp. $s[B/j]$) is obtained from the move
$m$ (resp. the play $s$) by replacing each token of the form $\star^A$
by $\star^{A[B/X_j]}$. Note that $s[B/j]$ is a play, but does not
necessarily belong to any $\MM_A$ for some $A$: actually, this play
will only be used as an intermediate construction.

\begin{Def}[copycat extension of a typed play]
Let $s=m_1\dots m_n$ be a typed play on the game $A$, let $B\in\mc{G}$
and $j>0$.

We first define the \textbf{flat extension} of $s$: given a sequence
of initial moves $r=(r_i)_{i\in\mathbb{N}}$ in $\MM_B$,
$\textit{Fl}^s_{j,B}(r)$ is the play $t[B/j]$ where $t$ is obtained
from $s$ by replacing each sequence $m_im_{i+1}$ such that $\sharp(m_i)=j$
and $\lambda(m_i)=\OO$ by $m_i[r_i]m_{i+1}[r_i]$.

Let $m_i$ be an $\OO$-move of $s$ such that $\sharp(m_i)=j$, suppose
$\textit{Fl}^s_{j,B}(r)=s_1m'_i[r_i]m'_{i+1}[r_i]s_2$ with $m'_i=m_i[B/j]$ and
$m'_{i+1}=m_{i+1}[B/j]$, and let $v=n_1\dots n_p\in\mc{BV}(B)$. The
\textbf{$B$-copycat extension} of $s$ at position $i$ along the index $j$ (with
parameters $v,r$) is the play $s'=CC^s_{j,B}(i,v,r)$ defined by:
\begin{itemize}
\item $s'=s_1m'_{i}[n_1]m'_{i+1}[n_1]s_2$
if $p=1$
\item $s'=s_1m'_{i}[n_1]m'_{i+1}[n_1]m'_{{i}+1}[n_2]m'_{i}[n_2]\dots
m'_{{i}+1}[n_p]m'_{i}[n_p]$ if $p$ even
\item $s'=s_1m'_{i}[n_1]m'_{i+1}[n_1]m'_{{i}+1}[n_2]m'_{i}[n_2]\dots
m'_{i}[n_p]m'_{i+1}[n_p]$ if $p>1$ and $p$ odd
\end{itemize}
\end{Def}

\begin{Def}[symbolic strategy]
A play $s$ on the game $A$ is said to be \textbf{symbolic} if,
whenever a game is played by $\OO$ it is a variable game
$X_i\notin FTV(A)$. These variable games are called the \textbf{copycat variables} of
the play.

A symbolic strategy is a strategy which contains only symbolic plays.
\end{Def}

\begin{Def}[copycat extension of an innocent symbolic strategy]
The copycat extension of an innocent symbolic strategy $\bar{\sigma}:A$ is
the smallest innocent strategy which contains $\barsig$
and is stable under any copycat extension along a copycat variable.
\end{Def}

\begin{Def}[uniform strategy]
Let $\sigma$ be a strategy on the game $A$. $\sigma$ is said to be
\textit{uniform} if there exists a symbolic innocent strategy
$\bar{\sigma}$ on $A$ such that $\sigma$ is the copycat extension of
$\bar{\sigma}$.
\end{Def}


\begin{Prop}\label{compunif}
If $\sigma::A\fl B$ and $\tau::B\fl C$ are two uniform strategies then
$\sigma;\tau::A\fl C$ is uniform.
\end{Prop}

The proof of this proposition can be found in appendix~\ref{proofcomp}.
\label{ftyped}

\section{The Curry-style model}\label{dcurry}





We are now ready to define our model: the key ingredient will be to
relate untyped strategies with typed strategies through a notion of
\textbf{realization}. First we relate untyped moves with typed moves
through an operation of \textbf{erasure}
$\erase:\mathbb{M}\fl\mathbb{X}$ defined by:
$$\erase=\mc{E}\circ\mc{A}$$


\begin{Def}[realization]
Let $\sigma$ be an untyped strategy and $\tilde{\sigma}$ a typed
strategy on $A$. We say that $\tilde{\sigma}$ is a
\textbf{realization} of $\sigma$ on $A$ if we have: for every
$sxy\in\sigma$ and $s'\in\tilde{\sigma}$, if
$s'm'\in\mc{P}_A$ is such that $\erase(s'm')=sx$ then there exists
$n'$ such that $s'm'n'\in\tilde{\sigma}$ and $\erase(s'm'n')=sxy$.
\end{Def}


\medskip

At present we have all the ingredients to define the model:
\begin{itemize}
\item \textbf{objects} are games
\item a \textbf{morphism} between $A$ and $B$ is an untyped strategy
  $\sigma$ such that:
\begin{itemize}
\item $\sigma$ is hyperuniform
\item there exists a typed strategy $\tilsig$ which is a realization
  of $\sigma$ on $A\fl B$
\item $\tilsig$ is uniform.
\end{itemize}
In this case we note $\sigma:A\fl B$.
\end{itemize}

\bigskip

Let us prove that we have a model of Curry-style system F indeed.


\begin{Lemme}
If $\sigma:A\fl B$ and $\tau:B\fl C$ then $\sigma;\tau:A\fl C$.
\end{Lemme}

\proof If we note $\tilsig$ and $\tiltau$ two realizations of $\sigma$
and $\tau$ respectively, we obtain a realization of $\sigma;\tau$ on
$A\fl C$ by taking the composite $\tilsig;\tiltau$ \textit{in the
grammar $\mathbb{M}$}. Indeed, suppose $sxy\in\sigma;\tau$,
$s'\in\tilsig;\tiltau$ with $\erase(s')=s$ and $s'm'\in\mc{P}_{A\fl
C}$ with $\erase(s'm')=sx$.  There exist an untyped justified
sequence $u$ such that $u\restrun=s_1\in\sigma$, $u\restrdeux=s_2\in\tau$ and
$u\restrtrois=sxy$, and a typed justified sequence $t$ such that
$t\restrun=t_1\in\tilsig$, $t\restrdeux=t_2\in\tiltau$ and $t\restrtrois=s'$.

We note $u=u_0xb_1\dots b_qy$, with $b_1,\dots b_q$ of shape
$\dna\upa$.  Suppose for example that $m'$ is of shape $\dna$. Then
there exists $n'_1$ such that $t_1m'n'_1\in\tilsig$ and
$\erase(t_1m'n'_1)=s_1xb_1$; we set $T_1=tm'n'_1$. Then there exists
$n'_2$ such that $t_2n'_1n'_2\in\tiltau$\footnote{More precisely
$n'_1=\upa n''$ should be renamed as $\dna n''$.} and
$\erase(t_2n'_1n'_2)=s_2b_1b_2$; we set $T_2=tm'n'_1n'_2$, etc. So, we
construct step by step a justified sequence $T$ such that
$T\restrun\in\tilsig$, $T\restrdeux\in\tiltau$ and $\erase(T)=u$. This
gives us also that $T\restrtrois=s'm'n'$ is a play, so it belongs to
$\tilsig;\tiltau$ and $\erase(s'm'n')=sxy$.


Finally: $\tilsig$ and $\tiltau$ are innocent and uniform, so
$\tilsig;\tiltau$ is innocent and uniform by prop.~\ref{compunif}; $\sigma$ and
$\tau$ are hyperuniform so $\sigma;\tau$ is hyperuniform by lemma~\ref{hyperunif2}.\cqfd

\begin{Lemme}
If $\sigma:\Gamma\fl A$ and $X_j\notin\Gamma$ then $\sigma:\Gamma\fl
\forall X_j.A$
\end{Lemme}

\proof Let us consider $\tilde{\sigma}::\Gamma\fl A$ a realization of
$\sigma$ on $\Gamma\fl A$: if $\tilsig$ is the copycat extension of a
symbolic strategy $\barsig$, then we define the strategy $\barsig'$ as
the strategy $\barsig$ where each move written $\upa m$ in a play has
been replaced by $\upa\star^{X_j}m$. This strategy is symbolic on
$\Gamma\fl \forall X_j.A$, and its copycat extension $\tilsig'$ is a
realization of $\sigma$ because of hyperuniformity (indeed, the only
difference between $\tilsig$ and $\tilsig'$ is a copycat extension
along $X_j$).\cqfd

\begin{Lemme}
If $\sigma:\Gamma\fl\forall X_j.A$ and $B$ is a game then
$\sigma:\Gamma\fl A[B/X_j]$.
\end{Lemme}

\proof If $\tilsig$ is a realization of $\sigma$ on $\Gamma\fl\forall
X_j.A$, a realization $\tilsig'$ on $\Gamma\fl A[B/X_j]$ is obtained
by taking only plays where each initial move takes the form
$\upa\star^Bm$, and by replacing each move $\upa\star^Bm$ by $\upa m$.

Let us now prove the uniformity of $\tilsig'$: if $\tilsig$ is the
copycat extension of a symbolic strategy $\barsig$, we consider a view
$s$ of $\barsig$. Let $X_j$ be the first copycat variable appearing in
$s$, we choose a variable $X_k\notin FTV(A)\cup FTV(B)$ and we call
$s_k$ the (unique) $X_k$-copycat extension of $s$ along $j$. Let us
define $E(s)$ as the smallest set of plays containing $s_k$ and stable
by $B$-copycat extensions along $k$. The strategy $\barsig'$ will be the
smallest innocent strategy containing all the sets $E(s)$, for $s$
describing all the views of $\barsig$. Then one can check that
$\tilsig'$ is the copycat extension of $\barsig'$.  \cqfd


\begin{Lemme} The following holds:
\begin{itemize}
\item $id:A\fl A$
\item $\pi_r:\Gamma\times A\fl A$
\item If $\sigma:\Gamma\fl A$ and $\tau:\Gamma\fl B$ then $\langle\sigma,\tau\rangle:\Gamma\fl(A\times B)$.
\item $eval:(A\fl B)\times A\fl B$
\item If $\sigma:\Gamma\times A\fl B$ then $\Lambda(\sigma):\Gamma\fl(A\fl B)$.
\end{itemize}
\end{Lemme}




These cases are trivial: for example, a realization of $\id$ on $A\fl
A$ is $$\rho=\{s\in\mc{P}_{A\fl A}\mid\ s\text{ of arrow shape }\text{ and }\forall
  t\in\mathbb{E},t\preceq s\Rightarrow t\restr_\upa=t\restr_{\dna}\}$$
  and it is uniform, with symbolic strategy $\bar{\rho}$ defined by:
$$\bar{\rho}=\{s\in\mc{P}_{A\fl A}\mid\ s\text{ of arrow shape, }s\text{
  symbolic }\text{ and }\forall
  t\in\mathbb{E},t\preceq s\Rightarrow t\restr_\upa=t\restr_{\dna}\}$$

\bigskip

If $\Gamma$ is a typing context of the form
$\Gamma=x_1:A_1,x_2:A_2,\dots,x_n:A_n$, we define the sequence of variables
$\overline{\Gamma}=x_1,x_2,\dots,x_n$ and the type $|\Gamma|=A_1\times
A_2\times\dots\times A_n$, and we have:
\begin{Prop}
If $\Gamma\vdash t:A$ then $\trad{\overline{\Gamma}\vdash t}:|\Gamma|\fl A$.
\end{Prop}

This, together with prop.~\ref{jul}, means that we have obtained a
model of Curry-style system F.\label{fcurry}

\section{Hyperforests}\label{hypf}


In this section we introduce the notion of \textbf{hyperforest}, an
arborescent structure built from games. In ~\cite{lataillade},
following~\cite{phdhughes}, we interpreted second-order types directly
as hyperforests (that we called polymorphic arenas). But the
substitution was difficult to define in this context, and moves had a
complicated formulation; that is why in this paper we introduce
hyperforests only as an indirect interpretation of types.

Hyperforests will be the fundamental structure for our work on
isomorphisms.

\subsection{Forests and hyperforests}

In what follows, the set of subsets of a set $E$ will be denoted $\mathbb{P}(E)$.

\begin{Def}[forest]
A \textbf{forest} is an ordered set $(E,\leq)$ such that, for every $y$ in $E$,
$\{x\mid x\leq y\}$ is finite and totally ordered by $\leq$. The forest is \textbf{finite}
if $E$ is finite.
\end{Def}

\begin{Def}[hyperforest]
An \textbf{hyperforest} $H=(\mc{F},\mc{R},\mc{D})$ is a finite
forest $\mc{F}$ together with a set of \textbf{hyperedges}
$\mc{R}\subseteq \mc{F}\times \mathbb{P}(\mc{F})$ and a partial
function of \textbf{decoration} $\mc{D}:\mc{F}\rightharpoonup \mc{X}$, where:
\begin{itemize}
\item for every $(t,S)\in\mc{R}$, if $s\in S$ then $t\leq s$ and
  $\mc{D}(s)$ is undefined
\item for every $b=(t,S)$ and $b'=(t',S')$ in $\mc{R}$, 
  $S\cap S'\neq\emptyset\Rightarrow b = b'$
\end{itemize}

We note $\mc{T}^H=\{t\in\FF\mid\ \exists S\subseteq\FF,\ (t,S)\in\RRR\}$ and
$\mc{S}^H=\{s\in\FF\mid\ \exists (t,S)\in\RRR,\ s\in S\}$.
\end{Def}

\begin{Def}[reference, friends]
Let $H=(\mc{F},\mc{R},\mc{D})$ be an hyperforest. For any $s\in\mc{F}$, if $s\in \mc{S}^H$
then there exists $(t,S)\in\RRR$ with $s\in S$: the \textbf{reference}
of $s$ is defined as $\rref^H(s)=t$ and the set of \textbf{friends} of
$s$ is $\friends^H(s)=S\backslash \{s\}$. If $s\notin \mc{S}^H$,
$\rref^H$ and $\friends^H$ are not defined in $s$.
\end{Def}

We are now going to exhibit the hyperforest structure
associated with a game $A$.


\subsection{From partially ordered sets to forests}

Let $(E,\leq)$ be a partially ordered set. The relation
$\vdash\subseteq E\cup(E\times E)$ is
given by: $$\begin{cases}\vdash e&\text{iff $e'\leq e\Rightarrow
    (e'=e)$}\\
e\vdash e'&\text{iff $e\leq e'\wedge\forall f,\ e\leq f\leq
e'\Rightarrow (e=f\vee e'=f)$}\end{cases}$$

One defines the set $F$ of \textbf{paths} in $(E,\leq)$, \ie\ the set
of sequences $e_1e_2\dots e_n$ of elements of $E$ such that $\vdash
e_1$ and $e_i\vdash e_{i+1}$ for $1\leq i\leq n-1$. If we consider the
prefix ordering $\leq'$ on $F$, then $(F,\leq')$ is a forest.

We also define the operation $\orig:F\fl E$ by $\orig(f)=e_n$ if $f=e_1\dots
e_n$ ($\orig(f)$ is called the \textbf{origin} of $f$).

\subsection{From games to hyperforests}


If $A$ is a game, $\occ_A$ is a finite partially ordered set, to which
one can associate a forest $\mc{F}_A$ through the preceding
construction. Extending $\vdash$ to $\mc{F}_A$ generates the enabling
relation of the forest: this justifies \textit{a posteriori} the
definition of an enabling relation for arbitrary moves given in
section~\ref{general}.

Furthermore, one deduces from $\LLL_A$ the relation
$\RRR_A\subseteq\mc{F}_A\times\mathbb{P}(\mc{F}_A)$ as follows: let
$\mc{L}=\{a[\star 0]\in\mathbb{A}\mid\ \exists a'\in\occ_A,a[\star 0]\sqsubseteq^p
a'\}$. Then :
\bec $(t,S)\in\RRR_A$ iff there exists $y\in\mc{L}$ such that, for every
$s\in S$:\eec
\begin{minilist}
\item $\LLL_A(\orig(s))=y$
\item $t\leq s$
\item $y\sqsubseteq^p\orig(t)$
\item for every $t'\leq t$, $y\sqsubseteq^p\orig(t')$ implies $t'=t$.
\end{minilist}

\medskip

One also defines the partial function $\DD_A:\mc{F}_A\rightharpoonup \mc{X}$ by:
$\DD_A(x)=X_i$ iff $\sharp(\orig(x))=i$ ($i>0$).

\medskip

Then we have:
\begin{Lemme}
If $A$ is a game, then $H_A=(\mc{F}_A,\RRR_A,\DD_A)$ is an hyperforest.
\end{Lemme}

\medskip

\paragraph{Example:}
Consider the type $A=\forall X_1.((X_1\times X_2)\fl (X_1\times \bot))$. We have:
$$\occ_{A}=\{\star\dna l0,\star\dna r2,\star\upa l0,\star\upa r0\}$$
and:
$$\begin{cases}
\begin{array}{lcc}
\mc{L}_A(\star\dna l0)&=&\star0\\
\mc{L}_A(\star\dna r2)&=&\dag\\
\mc{L}_A(\star\upa l0)&=&\star0\\
\mc{L}_A(\star\upa r0)&=&\dag
\end{array}&
\end{cases}$$
The paths are: $a=\star\upa l0$, $ b=\star\upa l0\cdot\star\dna
l0$, $ c=\star\upa l0\cdot\star\dna r2$, $ d=\star\upa r0$, $
e=\star\upa r0\cdot\star\dna l0$ and $ f=\star\upa r0\cdot\star\dna
r2$. Besides, $\mc{L}=\{\star 0\}$.

Hence the hyperforest $H_A$ is given by:
$$\mc{F}_A=\{a,b,c,d,e,f\}$$
$$\RRR_A=\{(a,\{a,b\}),(d,\{e\})\}$$
$$\DD_A(c)=\DD_A(f)=X_2$$

This can be resume in the following representation of $H_A$:

\smallskip

\begin{center}
\epsfig{file=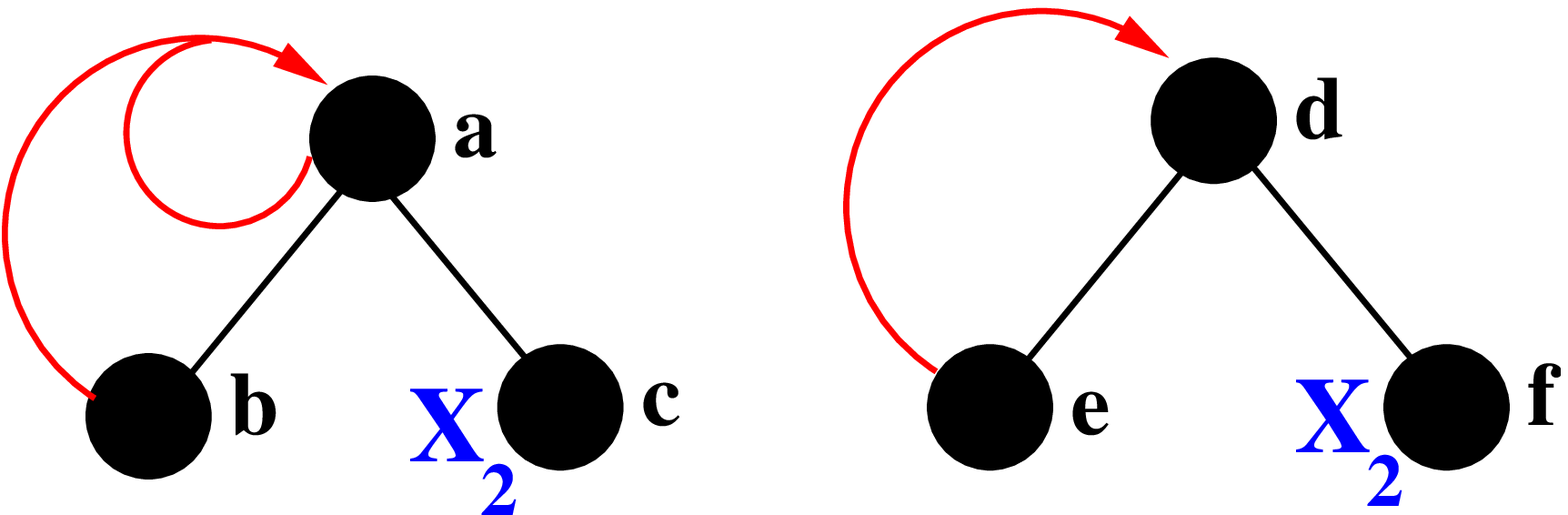,height=2cm}
\end{center}

\bigskip
\bigskip

One can extend the definition of polarity to the nodes of the
hyperforest: if $A$ is a game with associated hyperforest
$H_A=(\mc{F}_A,\RRR_A,\DD_A)$, then for $a\in\mc{F}_A$ we define
$\lambda(a)=\lambda(\orig(a))$. This coincides with an alternative
definition of polarity, which is common in arena games: $\lambda(a)=\OO$ (resp. $\lambda(a)=\PP$) if the set
$\{a'\in\mc{F}_A\mid\ a'\leq a\}$ has an odd cardinality (resp. an
even cardinality). Note also that
$\paux_A(\orig(a))=\lambda(\rref_A(a))$.

Finally, if $A$ is a game, we note:
$$\friends_A=\friends^{H_A}\qquad \rref_A=\rref^{H_A}\qquad \mc{S}_A=\mc{S}^{H_A}\qquad \mc{T}_A=\mc{T}^{H_A}$$

\medskip

Note that the nodes of the forest $\FF_A$ contain ``more
information'' than the occurrences of $\occ_A$. Indeed, given a node
$c\in\mc{F}_A$, one is able to give the ordered list of its ancestors,
whereas for an occurrence we may have many ancestors that are not
compatible one with the order for the ordering. This idea will be used in the proof of
theorem~\ref{main} to reason about plays with nodes instead of
occurrences.\label{fhyp}

\section{Type isomorphisms}

\subsection{Isomorphisms in the model}

\begin{Def}[Church-isomorphism]
Let $H_1=(\mc{F}_1,\RRR_1,\DD_1)$ and $H_2=(\mc{F}_2,\RRR_2,\DD_2)$ be
two hyperforests.  We say that $H_1$ and $H_2$ are
\textbf{Church-isomorphic} ($H_1\simeq_{\textit{Ch}} H_2$) if there
exists a bijection $f:\mc{F}_1\fl\mc{F}_2$ which preserves the
hyperforest structure, \ie\ such that:
\begin{itemize}
\item $a\leq a'$ iff $f(a)\leq f(a')$
\item $\RRR_2=f(\RRR_1)$
\item $\DD_2\circ f=\DD_1$
\end{itemize}
\end{Def}

\begin{Def}[Curry-isomorphism]
Let $H_1=(\mc{F}_1,\RRR_1,\DD_1)$ and $H_2=(\mc{F}_2,\RRR_2,\DD_2)$ be
two hyperforests.  We say that $H_1$ and $H_2$ are
\textbf{Curry-isomorphic} ($H_1\simeq_{\textit{Cu}} H_2$) if there exists a bijection
$f:\mc{F}_1\fl\mc{F}_2$ such that:
\begin{itemize}
\item $a\leq a'$ iff $f(a)\leq f(a')$
\item $\mc{S}^{H_2}=f(\mc{S}^{H_1})$
\item for every $(t,S)\in\RRR_1$ (resp. $(t,S)\in\RRR_2$), 
if there exists $s\in S$ such that $\lambda(s)\neq\lambda(t)$, then $(f(t),f(S))\in\RRR_2$ (resp. $(f^{-1}(t),f^{-1}(S))\in\RRR_1$)
\item $\mc{D}_2\circ f=\mc{D}_1$.
\end{itemize}
\end{Def}

\medskip

\begin{Def}[game isomorphism]
A \textbf{game isomorphism} between two games $A$ and $B$ is a couple of untyped strategies
$\sigma:A\fl B$ and $\tau:B\fl A$ such that
$\sigma;\tau=\tau;\sigma=\id$. We note $A\simeq_gB$ if there is a game
isomorphism between $A$ and $B$.
\end{Def}

\bigskip

We are now able to formulate the key theorem of our paper. This
theorem provides a geometrical characterisation of isomorphisms in the
model, which is the core of the proof of equational characterisation for the
syntax.

\begin{Th}\label{main}
Let $A,B\in\mc{G}$. If there exists a game isomorphism $(\sigma,\tau)$
between $A$ and $B$ ( $A\simeq_gB$) then their hyperforests are
Curry-isomorphic ($H_A\simeq_{\textit{Cu}}H_B$).
\end{Th}

The proof of this theorem can be found in appendix~\ref{proofmain}.

\subsection{Characterisation of Curry-style type isomorphisms}

Proving theorem~\ref{main} was the main step towards the
characterisation of Curry-style isomorphisms: we are now able to
establish our final result.

Let us recall the equational system $\eq$ which we claim to
characterise Curry-style type isomorphisms:
\begin{align*}
A\times B&\eq B\times A&\\
A\times (B\times C)&\eq(A\times B)\times C&\\
A\fl (B\fl C)&\eq(A\times B)\fl C&\\
A\fl (B\times C)&\eq(A\fl B)\times(A\fl C)&\\
\forall X.\forall Y.A&\eq\forall Y.\forall X.A&\\
A\fl \forall X.B&\eq\forall X.(A\fl B)&\text{if $X\notin FTV(A)$}\\
\forall X.(A\times B)&\eq\forall X.A\times\forall X.B&\\
\forall X.A&\eq A[\forall Y.Y/X]&\text{if $X\notin\negocc_A$}
\end{align*}

\begin{Lemme}\label{eqsyst}
Let $A$ and $B$ be two types such that the hyperforests
$H_A$ and $H_B$ are Curry-isomorphic.
Then $A$ and $B$ are equal up to the equational system $\eq$.
\end{Lemme}

\proof Let $A'$ and $B'$ be the normal forms of $A$ and $B$ for the
following rewriting system:
$$\forall X.C\reec C[\forall Y.Y/X]\quad\text{ if $X\notin\negocc_C$
and $C\neq X$}$$

If $D_1=\forall X.C$ and $D_2=C[\forall Y.Y/X]$ with
$X\notin\negocc_C$, then $H_{D_1}\simeq_{\textit{Cu}} H_{D_2}$:
indeed, the bijection $f:\mc{F}_{D_1}\fl\mc{F}_{D_2}$ which preserves
the ordering and such that $\mc{S}_{D_2}=f(\mc{S}_{D_1})$ and
$\mc{D}_{D_2}\circ f=\mc{D}_1$ is easy to define (in fact
$\occ_{D_1}$ and $\occ_{D_2}$ are already in
bijection). The fact that $X\notin\negocc_C$ precisely implies that,
for any $(t,S)\in\RRR_{D_1}$ corresponding to the quantification
$\forall X$ (\ie\ such that $\mc{L}_{\forall X.A}(\orig(s))=\star0$ for
every $s\in S$), there is no $s\in S$ such that
$\lambda(s)\neq\lambda(t)$. Reciprocally, if for any
$(t,S)\in\RRR_{D_2}$ corresponding to a quantification $\forall Y.Y$,
$S=\{t\}$ so there is no $s\in S$ such that
$\lambda(s)\neq\lambda(t)$. Any other hyperedge is preserved by $f$.

Moreover, being Curry-isomorphic is a congruence (\ie\ it is preserved
by context), so $H_A\simeq_{\textit{Cu}} H_{A'}$,
$H_B\simeq_{\textit{Cu}} H_{B'}$, and hence
$H_{A'}\simeq_{\textit{Cu}} H_{B'}$. $H_{A'}$ and $H_{B'}$ are such
that for every $(t,S)\in\RRR_{A'}$ (or $(t,S)\in\RRR_{B'}$), either
$S=\{t\}$ or $S$ contains a node $s$ with
$\lambda(t)\neq\lambda(s)$. Because of the definitions of
$\simeq_{\textit{Cu}}$ and $\simeq_{\textit{Ch}}$, this implies $H_{A'}\simeq_{\textit{Ch}}
H_{B'}$.

It has already been proved
in~\cite{lataillade}\footnote{In~\cite{lataillade} the interpretation
of types was directly hyperforests.}
that in this case $A'\eq' B'$, where $\eq'$ is the same equational
system as $\eq$, except that it does not make use of the last
equation. Hence, we have $A\eq B$.  \cqfd

\medskip

\begin{Th}\label{isoscurry}\label{charact}
Two types $A$ and $B$ are isomorphic in Curry-style system F if and only if $A\eq B$.
\end{Th}

\proof The implication comes from the fact that we have a model (so,
each type isomorphism in Curry-style system F implies a game
isomorphism) and from theorem~\ref{main} and lemma~\ref{eqsyst}.

For the reciprocal, we already know from~\cite{isotypes} the existence
in the Church-style system F of the isomorphisms corresponding to each
equation of $\eq$, except the last one ($\forall X.A\eq A[\forall
  Y.Y/X]\text{ if }X\notin\negocc_A$). This implies their existence
in the Curry-style system F.

Hence, we need, given a type $A$ such that $X\notin\negocc_A$, to find
two Curry-style terms $t:\forall X.A\fl A[\forall Y.Y/X]$ and
$u:A[\forall Y.Y/X]\fl\forall X.A$ which compose in both ways to give
the identity. We suppose $Y$ does not appear at all in $A$, even as a
bounded variable.

We take $t=\lambda x.x$: indeed, the identity can be shown to
be of type $\forall X.A\fl A[\forall Y.Y/X]$ through the following
type derivation:
\begin{prooftree}
\AxiomC{$x:\forall X.A\vdash x:\forall X.A$}
\UnaryInfC{$x:\forall X.A\vdash x:A[\forall Y.Y/X]$}
\UnaryInfC{$\vdash\lambda x.x:\forall X.A\fl A[\forall Y.Y/X]$}
\end{prooftree}

$t$ is easy to build: consider the Church-style term
$M=\lambda x^{\forall X.A}. (x)\{\forall Y.Y\}$. We have $\vdash
M:\forall X.A\fl A[\forall Y.Y/X]$ in Church-style system F, and $t$
is the $\lambda$-term obtained by erasing each type indication in
$M$. Then we necessarily have $\vdash t:\forall X.A\fl A[\forall
Y.Y/X]$, and besides $t=\lambda x.x$.

To define $u$, let us consider the Church-style term $P$ which is the
$\eta$-long normal form of the identity on $A[\forall Y.Y/X]$. This term takes the
form $P=\lambda x^{A[\forall Y.Y/X]}.P'$. Now consider the Church-style term $Q$
obtained from $P'$ by replacing each occurrence of $y\{Z\}$, where $Z$
is some type variable and $y$ has the type $\forall Y.Y$ coming
from the substitution of $X$, by $y\{X\}$. For
example, if $A=X\fl\bot\fl\bot$, this would give us $Q=(x)\lambda
y^{(\forall Y.Y)\fl\bot}.(y)\lambda z^{\forall Y.Y}.(z)\{X\}$

Then we introduce the Church-style term $N=\lambda x^{A[\forall
    Y.Y/X]}.\Lambda X.Q$, and we can check that $\vdash N:A[\forall
  Y.Y/X]\fl\forall X.A$ in Church-style system F. $u$ is now defined
to be the erasure of $N$. Then we necessarily have $\vdash u:A[\forall
  Y.Y/X]\fl\forall X.A$, and besides $u=\lambda x.x$ (modulo
$\eta$-reductions) because we only modified the type indications when
going from $P$ to $N$.

Finally, $t$ and $u$ trivially compose to give the identity in
both directions. 
\cqfd


\section*{Conclusion}

We have proved that type isomorphisms in Curry-style system F can be
characterised by adding to the equational system of Church-style
system F isomorphisms a new, non-trivial equation: $\forall X.A\eq
A[\forall Y.Y/X]$ if $X\notin\negocc_A$. Otherwise said, this equation
characterises all the new type equivalences one can generate by erasing
type indications in Church-style terms.

We used a game semantics model in order to take advantage of its
dynamical and geometrical properties. The main features of the model
were however often inspired by a precise analysis of the syntax:
indeed, an interpretation of the quantifier as an intersection (or a
lower bound like in~\cite{phdjulius}) was not precise enough to be
able to characterise type isomorphisms.

One can notice that our type system does not contain the type $\top$;
correspondingly, our model has no empty game. This is because the rule
generally associated to $\top$ takes the form: $t=\star$ if
$\Gamma\vdash t:\top$. This rule is of course difficult to insert in a
Curry-style setting, where terms are not typed a priori, and we have
no clue whether such a rule can be adapted to this context. Anyway,
the introduction of an empty game in the model would break the proof
and, more interestingly, give raise to new isomorphisms like $\forall
X.(X\fl\bot)\simeq_g\bot$. The characterisation of isomorphisms in this
model, and the possible connection with an actual syntax, have to be
explored.

But the main trail of future exploration concerns parametric
polymorphism. The notion of relational parametricity, introduced by
Reynolds~\cite{reynolds83}, comes historically from the idea that a
second-order function shall not depend on the type at which it is
instantiated. This has led first to a semantic definition of
parametricity, then to a syntactic formalisation of this notion, first
by Abadi-Cardelli-Curien~\cite{param} and then by
Plotkin-Abadi~\cite{paramlogic}. Dunphy~\cite{dunphy} recently gave a
categorical characterisation of parametric polymorphism.

The great advantage of parametric models is that second-order enjoys
nice and natural properties in these models. For example:
\begin{itemize}
\item $\forall X.X\fl X$ is a terminal object
\item $\forall X.(A\fl B\fl X)\fl X$ is a product of $A$ and $B$
\item $\forall X.X$ is an initial object
\item $\forall X.(A\fl X)\fl(B\fl X)\fl X$ is a coproduct of $A$ and $B$.
\end{itemize}
All these properties are of course wrong in the model described in the
present paper.

Trying to build a parametric game model is a highly appealing
challenge: one would be glad to extend the concrete notions and
flexible features of games into a context where parametricity is
understood. Studying isomorphisms in this context would be a natural
question, considering the particularly powerful ones corresponding to
the above properties.

Finally, relational parametricity seems to be related to Curry-style
system F, if we believe in a conjecture of Abadi-Cardelli-Curien which
says the following: suppose you have two terms of type A whose type
erasures are the same. Then they are parametrically equal (the
converse is false). This means that the parametric equality is
(strictly) stronger than the Curry-style equality: the study on both
Curry-style system F and parametricity in the context of games may
help to explore this question.

\bibliographystyle{alpha}
\bibliography{ll}

\newpage

\appendix

\setcounter{Prop}{1}

\section{Uniform strategies compose}
\label{proofcomp}

\begin{Prop}
If $\sigma::A\fl B$ and $\tau::B\fl C$ are two uniform strategies then
$\sigma;\tau::A\fl C$ is uniform.
\end{Prop}

\proof
Consider the following strategy
$$
\bar{\rho}=\{u\restrtrois\mid\ u\in\textbf{Int}\wedge
u\restrun\in\sigma\wedge  u\restrdeux\in\tau
\wedge u\restrtrois\text{ symbolic
  play}\}$$
It is an innocent strategy on $A\fl C$ (the proof is the same as in HON models),
and it is of course symbolic. We call $\rho$ its
copycat extension, and we want to prove that $\rho=\sigma;\tau$.

First we prove that $\rho\subseteq\sigma;\tau$: as
$\bar{\rho}\subseteq\sigma;\tau$, we need to show that $\sigma;\tau$
is stable by any copycat extension along any index $j$.  Note that, if
the variable game $X_j$ is played by $\OO$ in $A\fl C$, it is also
played by $\OO$ in $A\fl B$ or $B\fl C$.  Consider the play
$s'=\textit{Fl}^s_{j,D}(r)$ for $s=m_1\dots m_n\in\sigma;\tau$,
$D\in\mc{G}$ and $r$ sequence of initial move in $\MM_D$. One shows
that $s'\in\sigma;\tau$: indeed there exist a justified sequence $u$
and two plays $s_1\in\sigma$ and $s_2\in\tau$ such that
$u\restr_{\dna\upa,\dna\dna}=s_1$, $u\restr_{\upa,\dna\upa}=s_2$ and
$u\restr_{\upa,\dna\dna}=s$.  Let us consider the justified sequence
$U_0$ obtained from $u$ by replacing each sequence $m_ib_1\dots
b_qm_{i+1}$ by $m_i[r_i]b_1[r_i]\dots b_q[r_i]m_{i+1}[r_i]$, and set
$U=U_0[D/j]$.  Then $U\restrun=s'_1\in\sigma$ (it is a flat extension,
hence a copycat extension, of $s_1$), $U\restrdeux\in\tau$ (it is a
flat extension, hence a copycat extension, of $s_2$) and
$U\restrtrois=s'\in\sigma;\tau$.

Now consider a move $m_i$ of $s$ such that $\sharp(m_i)=j$ and a bi-view
$v=n_1\dots n_p$ in the game $D$, and set $S=CC^s_{j,D}(i,v,r)$. If
$U=U_1m'_i[r_i]b_1[r_i]\dots b_q[r_i]m'_{i+1}[r_i]U_2$ with
$m'_i=m_i[D/j]$ and $m'_{i+1}=m_{i+1}[D/j]$, one can build another
justified sequence $U'$, depending on the value of $p$:
\begin{itemize}
\item if $p=1$, $U'=U_1m'_{i}[n_1]b_1[n_1]\dots b_q[n_1]m'_{i+1}[n_1]
U_2$
\item if $p$ even,\\
$U'=U_1m'_{i}[n_1]b_1[n_1]\dots b_q[n_1]m'_{i+1}[n_1]
m'_{i+1}[n_2]b_q[n_2]\dots b_1[n_2]m'_{i}[n_2]\dots\dots
m'_{i+1}[n_p]b_q[n_p]\dots b_1[n_p]m'_{i}[n_p]$
\item if $p$ odd and $p>1$,\\
$U'=U_1m'_{i}[n_1]b_1[n_1]\dots b_q[n_1]m'_{i+1}[n_1]
m'_{i+1}[n_2]b_q[n_2]\dots b_1[n_2]m'_{i}[n_2]\dots\dots
m'_{i}[n_p]b_1[n_p]\dots b_q[n_p]m'_{i+1}[n_p]$
\end{itemize}

$U'\restrun$ is a copycat extension of $s_1$ ($s'_1$ was the flat
extension) so $U'\restrun\in\sigma$, and similarly
$U'\restrdeux\in\tau$. $U'\restrtrois$ is a play so
$U'\restrtrois=S\in\sigma;\tau$.

\bigskip

The last thing to prove is that $\sigma;\tau\subseteq\rho$. We suppose
that $\sigma$ and $\tau$ are the copycat extensions of the symbolic
strategies $\barsig$ and $\bartau$ respectively.  Consider a play
$s\in\sigma;\tau$, there exists a justified sequence $u$ for which
$u\restr_{\dna\upa,\dna\dna}=s_1\in\sigma$,
$u\restr_{\upa,\dna\upa}=s_2\in\tau$ and
$u\restr_{\upa,\dna\dna}=s$.


Let $D_1,\dots,D_N$ be the sequence of games played by $\OO$ in $u$ at
the level of moves of shape $\upa$ or $\dna\dna$. Suppose for
simplicity that $X_1,\dots,X_N\notin FTV(A)$.  Consider a subsequence
$U=m[m_1]b_1[m_2]\dots b_q[m_{q+1}]n[m_{q+2}]$ of $u$ such that:
$\AAA(m),\AAA(b_1),\dots,\AAA(b_q),\AAA(n)\in\occ_{(A\fl B)\fl C}$,
$\paux_{(A\fl B)\fl C}(c)=\OO$ if $c=\AAA(m)$, and $b_1,\dots,b_q$ are
of shape $\dna\upa$ whereas $m,n$ are not of this shape. Suppose $U$
is the first such sequence in $u$ and $m$ is of shape $\upa$ (the case
where $m$ is of shape $\dna\dna$ is similar). Then
$\frac{m}{\LLL_{(A\fl B)\fl C}(c)}=D_j$ for some $1\leq j\leq N$ (it
is a game played by $\OO$, because $\paux_{(A\fl B)\fl C}(c)=\OO$).

If $u=u_1Uu_2$, we build a new sequence $u'$ as follows:
\begin{itemize}
\item $t_1=(u_1 m[m_1]b_1[m_2])\restrun\in\sigma$ is a $D_j$-copycat
  extension of some $\bar{s}_1\in\sigma$: indeed, $\sigma$ is the
  smallest innocent strategy that contains $\OO$ and is stable by
  copycat extension, so $t_1$ must be composed of many views that are
  obtained from $\barsig$ by copycat extensions; besides, $U$ is the
  first subsequence of its kind, so there is in fact only one of these
  copycat extensions that applies on a variable played at the level of
  a move of shape $\upa$ or $\dna\dna$ (so, only one
  $B_j$-extension). $\bar{s}_1$ takes the form $\bar{s}_1=(u'_1
  m[j]b_1[M_1])\restrun$ where $u'_1$ is obtained by replacing each
  occurrence of $D_j$ in $u_1$ by $X_{j}$
\item $t_2=(u_1 b_1[m_2]b_2[m_3])\restrdeux\in\tau$ is a $D_j$-copycat
  extension of some $\bar{s}_2\in\tau$: indeed, $\tau$ is the smallest
  innocent strategy that contains $\OO$ and is stable by copycat
  extension, so $t_2$ must be composed of many views that are obtained
  from $\bartau$ by copycat extensions; besides, $U$ is the first
  subsequence of its kind, so there is in fact only one of these
  copycat extensions that applies on a variable played at the level of
  a move of shape $\upa$ or $\dna\dna$ (so, only one
  $B_j$-extension). $\bar{s}_2$ takes the form $\bar{s}_2=(u'_1
  m[j]b_1[M_1])\restrdeux$
\item we iterate this process until we get to $n[M_{q+1}]$ for some
  $M_{q+1}$: this gives us a justified sequence $u'$ on $(A\fl B)\fl
  C$ which can be copycat extended to $u_1U$, and such that
  $u'\restrun\in\sigma$, $u'\restrdeux\in\tau$.
\end{itemize}
Now we iterate this process for each subsequence of $u$ having the
same properties as $U$, and what we obtain is a justified sequence $u''$ on
$(A\fl B)\fl C$ such that $u'\restrun\in\sigma$, $u'\restrdeux\in\tau$
and $t=u''\restrtrois$ is a play. Moreover each $D_j$ has been
replaced by $X_{j}$ (it might actually not be the case if some $D_j$ did
not correspond to any of our sequences, but in this case we just
replace $D_j$ by $X_{j}$ harmlessly), so $t\in\barrho$.

Finally, $u$ can be obtained from $u''$ by copycat extension, so $s$ can
be obtained from $t$ by copycat extension. Hence, $s\in\rho$.
\cqfd

\bigskip

\section{Proof of $A\simeq_gB\Rightarrow H_A\simeq_{\textit{Cu}}H_B$}
\label{proofmain}

\setcounter{Th}{0}

\begin{Def}[zig-zag play]
A play $s$ of arrow shape is said to be \textbf{zig-zag} if
\begin{minilist}
\item each Player move following an Opponent move of the form
$\upa m$ (resp.  $\dna m$) has the form $\dna m'$
(resp. $\upa m'$)
\item each (Player) move which follows an (Opponent) initial move is
justified by it
\item $s\restr_{\upa}$ and $s\restr_{\dna}$ have the same pointers.
\end{minilist}

If $s$ is a zig-zag even-length play, we note
$\breve{s}$ the unique zig-zag play such that
$\breve{s}\restr_\upa=s\restr_\dna$ and $\breve{s}\restr_\dna=s\restr_\upa$.
\end{Def}

\smallskip

\begin{Th}
Let $A,B\in\mc{G}$. If there exists a game isomorphism $(\sigma,\tau)$
between $A$ and $B$ ($A\simeq_gB$) then their hyperforests are
Curry-isomorphic ($H_A\simeq_{\textit{Cu}}H_B$).
\end{Th}

\bigskip

\textsc{Proof:} For the sake of simplicity, we will throughout this
proof identify the nodes of $\mc{F}_A$ (resp. of $\mc{F}_B$) with the
corresponding nodes of $\mc{F}_{A\fl B}$.

\medskip

\subsection*{Zig-zag property}


Let $\sigma:A\fl B$ and $\tau:B\fl A$ be the untyped strategies
which form the game isomorphism, and let $\tilde{\sigma}::A\fl B$
and $\tilde{\tau}::B\fl A$ be two realizations of $\sigma$ and $\tau$,
respectively on $A\fl B$ and $B\fl A$.

We begin with the following:
\begin{minilist}
\item every play of $\sigma$ or $\tau$ is zig-zag
\item $\tau=\{\breve{s}\mid\ s\in\sigma\}$
\item $\sigma$ and $\tau$ are total on the shape $\{\upa,\dna\}$.
\end{minilist}

This has been proven in a simply typed context, \ie\ with strategies playing
on forests, in~\cite{classisos}. The present situation is actually a
particular case of the simply typed one where the two forests to consider are
universal (in the sense that they contain any move). Totality for
universal forests immediately implies totality on the arrow shape.

One consequence of totality on the arrow shape is that, whenever
$s\in\tilsig$, we have $\erase(s)\in\sigma$.

\bigskip


\subsection*{Copycat property}

We now prove the following:
\begin{center}
if $sm_1[m'_1]m_2[m'_2]\in\tilsig$ with
$\AAA(m_1),\AAA(m_2)\in\occ_{A\fl B}$, then $\erase(m'_1)=\erase(m'_2)$
\end{center}

\medskip

We call it the \textbf{copycat property}.
Note that this property will hold only because $(\sigma,\tau)$ is a
game isomorphism, it is not true in general.

First consider the case where $S=sm_1[m'_1]m_2[m'_2]$ is symbolic.  We
note $u=\erase(S)$ and we have $u\in\sigma$ and $v=\breve{u}\in\tau$.
We will prove by recurrence that $\erase(m'_1)=\erase(m'_2)$, and that
it is possible to build a play $T\in\tiltau$ such that
$\erase(T)$ is a copycat extension of $v$. So, suppose this is true for every $t\in\mathbb{E}$
such that $t\preceq s$.

We set $\begin{cases}x_1=\erase(m_1)\\x_2=\erase(m_2)\end{cases}$,
$\begin{cases}a_1=\AAA(m_1)\\a_2=\AAA(m_2)\end{cases}$,
$\begin{cases}x'_1=\erase(m'_1)\\x'_2=\erase(m'_2)\end{cases}$ and
$\begin{cases}a'_1=\AAA(m'_1)\\b'_2=\AAA(m'_2)\end{cases}$

We have three cases:
\begin{itemize}
\item if $\paux_{A\fl B}(a_1)$ is undefined (so that $m'_1=0$),
  suppose $m'_2\neq 0$: then $\paux_{A\fl B}(a_2)=\PP$ (because
  $\sharp(m'_2)=0$ and $\OO$ has only played symbolically), so
  $\paux_{B\fl A}(a_2)=\OO$. As we have $T'\in\tiltau$ such that
  $t'=\erase(T')$ is a copycat extension of $t$ (so $t'\in\tau$ by
  hyperuniformity), one can build $T'm_2[j]\in\mc{P}_{B\fl A}$ for some
  $j\neq 0$ (remember $\OO$ plays symbolically), so by definition of
  the realization there exists $N\in\MM_{B\fl A}$ such that
  $T'm_2[j]N\in\tiltau$. Then $t'x_2[j]y\in\tau$ with $y=\erase(N)$
  and $\sharp(y)=j$, so by hyperuniformity
  $t'x_2[x'_2]y[x'_2]\in\tau$: this breaks determinacy since
  $tx_2[x'_2]x_1[0]\in\tau$ implies $t'x_2[x'_2]x_1[0]\in\tau$ by
  hyperuniformity, so $m'_2\neq 0$ is impossible. Finally, we have
  $T''=T'm_2[0]m_1[0]\in\tiltau$ with $\erase(T'')$ copycat extension
  of $v$.
\item if $\paux_{A\fl B}(a_1)=\OO$, then $sm_1[j]m_2[M[j]]\in\tilsig$
  for some $j$ such that $X_j\notin FTV(A)$, so $\paux_{A\fl B}(a_2)$
  is defined.  The case $\paux_{A\fl B}(a_2)=\PP$ implies $\paux_{B\fl
  A}(a_2)=\OO$; as we have $T'\in\tiltau$ such that $t'=\erase(T')$
  copycat extension of $t$ (so $t'\in\tau$ by hyperuniformity), there
  exists $k$ such that $T'm_2[k]\in\mc{P}_{B\fl A}$ so
  $T'm_2[k]N[N']\in\tiltau$ for some typed moves $N,N'$ with
  $\AAA(N)\in\occ_{B\fl A}$. Hence $t'x_2[k]y[y'[k]]\in\tau$ with
  $y=\erase(N)$ and $y'=\erase(N')$, and by hyperuniformity
  $t'x_2[x'_2]y[y'[x'_2]]\in\tau$.  But $tx_2[x'_2]x_1[j]\in\tau$
  implies $t'x_2[x'_2]x_1[j]\in\tau$ by hyperuniformity, so
  $y'[x'_2]=j$ (hence $x'_2=j$) and $y[j]=x_1[j]$.

The case $\paux_{A\fl B}(a_2)=\OO$ directly implies $M[j]=j$.  One
  still has to build in this case the play $T\in\tiltau$ such that
  $\erase(T)$ is a copycat extension of $v$: we have $T'\in\tiltau$ such
  that $t'=\erase(T')$ copycat extension of $t$ (so $t'\in\tau$ by
  hyperuniformity); moreover, if $D=\frac{m_2}{\LLL_A(a_2)}$ then
  there is at least one initial move $M[j]\in\MM_D$. So,
  $T'm_2[M[j]]\in\mc{P}_{B\fl A}$, and then $T=T'm_2[M[j]]N[N']\in\tiltau$
  for some typed moves $N,N'$ with $\AAA(N)\in\occ_{B\fl A}$.  Hence $t'x_2[z]y[y']\in\tau$ with
  $z=\erase(M)$, $y=\erase(N)$ and $y'=\erase(N')$. But
  $tx_2[j]x_1[j]\in\tau$ implies $t'x_2[z]x_1[z]$ by hyperuniformity,
  so by determinacy $\erase(T)=t'x_2[z]x_1[z]$: it is a copycat extension of $v$.
\item if $\paux_{A\fl B}(a_1)=\PP$, then $\paux_{B\fl A}(a_1)=\OO$. As
  we have $T'\in\tiltau$ such that $t'=\erase(T')$ copycat extension
  of $t$, one can build as
  above $T'm_2[M]\in\mc{P}_{B\fl A}$ for some typed move $M$ (if
  $D=\frac{m_2}{\LLL_A(a_2)}$ then there is at least one initial move
  $M\in\MM_D$), so $T'm_2[M]N[N']\in\tiltau$ for some $N,N'$ with
  $\AAA(N)=c\in\occ_{B\fl A}$. We set $y=\erase(M)$, $z=\erase(N)$ and
  $z'=\erase(N')$. There are two possibilities: if $\paux_{A\fl
  B}(a_2)=\OO$, then $M=j$ for some $j$. As $tx_2[j]x_1[x'_1]\in\tau$,
  one has $t'x_2[y]x_1[x'_1[y]]\in\tau$ by hyperuniformity, so
  $x_1[x'_1[y]]=z[z']$ by determinacy. This means $z[j]=x_1[j]$, so
  $\paux_{B\fl A}(c)=\OO$. Hence $z'[k]=k$, and so
  $y[k]=x'_1[k]=k$. If $\paux_{A\fl B}(a_2)=\PP$, then $\paux_{B\fl
  A}(a_2)=\OO$ so $y=j$ for some $j$. As $tx_2[k]z[z'[k]]\in\tau$, one
  has $t'x_2[x'_2]z[z'[x'_2]]\in\tau$ by hyperuniformity, so
  $z[z'[x'_2]]=x_1[x'_1]$ by determinacy. This means $z[k]=x_1[k]$, so
  $\paux_{B\fl A}(c)=\OO$. Then $z'[k]=k$, so $x'_1=x'_2$.
\end{itemize}

Finally, if $S$ is not symbolic, there exists a symbolic play
$S'=s'M_1[M'_1]M_2[M'_2]\in\tilsig$ whose $S$ is a copycat
extension. So $\erase(M'_1)=\erase(M'_2)$ and
$\erase(m'_1)=\erase(m'_2)$ because of the definition of the copycat
extension.

\bigskip

\subsection*{Construction of the untyped copycat play}


Let $a$ be a node of $\mc{F}_A$ and $a_1,\dots,a_p$ be the sequence of
nodes of $\mc{F}_A$ such that $\vdash a_1$, $a_i\vdash a_{i+1}$ and
$a_p=a$. We are going to construct a function $f:\mc{F}_A\fl\mc{F}_B$
such that, for any $i\in\mathbb{N}$:
$$\mc{E}'(f(a_1))[i]\mc{E}'(a_1)[i]\mc{E}'(a_2)[i]\mc{E}'(f(a_2))[i]\mc{E}'(f(a_3))[i]\mc{E}'(a_3)[i]\dots\in\sigma$$
where $\mc{E}'=\mc{E}\circ\orig:\mc{F}_A\cup\mc{F}_B\fl\mathbb{X}$.


The construction of $f$ will use the determinacy of $\sigma$ and
$\tau$, which generates a unique move starting from a play with its
complete history (not just the last move). That is why we could not
work with a function $f':\occ_A\fl\occ_B$, because in that case, the
choice of $f'(a)$ would depend not only on $a$, but also on the choice
of the ancestors. As said at the end of section~\ref{hypf}, the
information contained in a node $c\in\mc{F}_A$ is precisely the node
$\orig(c)\in\occ_A$ plus its ancestors: so, the forests are the good
structure to ensure that the function $f$ is well-defined. Having this
in mind, one can identify $\orig(a_i)$ (resp. $\orig(f(a_i))$) with
$a_i$ (resp. $f(a_i)$), and try to prove:
$\mc{E}(f(a_1))[i]\mc{E}(a_1)[i]\mc{E}(a_2)[i]\mc{E}(f(a_2))[i]\mc{E}(f(a_3))[i]\mc{E}(a_3)[i]\dots\in\sigma$.

Moreover, by the property of non-ambiguity (cf. def.~\ref{defgame}),
one has, for any $b$ in $\mc{F}_A$ (resp. in $\mc{F}_B$):
$\EEE(a_i)=\EEE(b)\Rightarrow b=a_i$
(resp. $\EEE(f(a_i))=\EEE(b)\Rightarrow b=f(a_i)$).
That is why we will also identify
$a_i$ (resp. $f(a_i)$) with $\EEE(a_i)$ (resp. $\EEE(f(a_i))$). What
has to be proved is then:
$$f(a_1)[i]a_1[i]a_2[i]f(a_2)[i]f(a_3)[i]a_3[i]\dots\in\sigma$$

\medskip

If $p=1$, we build a symbolic play $s=m_1\in\mc{P}_{B\fl A}$ such that
$\erase(m_1)=a_1[i]$ for $X_i=\frac{m_1}{\LLL_{B\fl A}(a_1)}$. Let $b$
be the only untyped move such that $a_1[i]b\in\tau$ (which exists by
totality of $\tau$). $\tiltau$ being a realisation of $\tau$, there
must be a play $m_1M\in\tiltau$ with $\erase(M)=b$, and we have a decomposition
$M=m_2[m'_2]$ with $\AAA(m_2)=c\in\occ_{B\fl A}$. We choose
$f(a_1)=c$, and we have $\erase(m'_2)=i$ because of the copycat
property.

\medskip

If $p=p'+1$ with $p'$ odd, we have by induction hypothesis:
$f(a_1)[i]a_1[i]a_2[i]f(a_2)[i]\dots f(a_{p'})[i]a_{p'}[i]\in\sigma$,
and by totality of $\sigma$ there exists a unique move $x$ such that
$f(a_1)[i]a_1[i]a_2[i]f(a_2)[i]\dots
f(a_{p'})[i]a_{p'}[i]a_{p}[i]x\in\sigma$. One is able to build
inductively a play $S\in\sigma$ such that
$\erase(S)=f(a_1)[y_1]a_1[y_1]\dots f(a_{p'})[y_{p'}]a_{p'}[y_{p'}]$
for a good choice of the moves $y_k$: indeed, if $S'\in\sigma$ with
$\erase(S')=f(a_1)[y_1]a_1[y_1]\dots f(a_{k})[y_{k}]a_{k}[y_{k}]$ (the
case $\erase(S')=f(a_1)[y_1]a_1[y_1]\dots a_{k}[y_{k}]f(a_{k})[y_{k}]$
is similar), we choose a move $M=m_1[m_2]$ with $\AAA(m_1)=a_{k+1}$
and $m_2$ initial move of $\MM_D$ where $D=\frac{m_1}{\LLL_{A\fl
B}(a_{k+1})}$; we note $\erase(m_{k+1})=y_{k+1}$. $\tilsig$ being a
realisation of $\sigma$, we have $S'MM'\in\tilsig$ for some typed move
$M'$, and $\erase(M')=a_{k+1}[y_{k+1}]$ by hyperuniformity of
$\sigma$. Hence we have obtained $S=S'MM'\in\tilsig$ such that
$\erase(S)=f(a_1)[y_1]a_1[y_1]\dots
a_{k+1}[y_{k+1}]f(a_{k+1})[y_{k+1}]$ for some $y_{k+1}\ $\footnote{In
the next part of the proof we will also build a typed play
$s_p\in\tilsig$ such that $\erase(s_p)=f(a_1)[y_1]a_1[y_1]\dots
f(a_{p})[y_{p}]a_{p}[y_{p}]$, but there will be more constraints on
$s_p$.}.

Then one chooses a typed move $N$ such that $tN\in\mc{P}_{A\fl B}$ and
$\erase(M)=a_p[y_p]$ for some initial move $y_p$ (it suffices once
again to choose $y_p$ as initial in $\MM_D$ for the appropriate game
$D$).

As $f(a_1)[y_1]a_1[y_1]\dots
f(a_{p'})[y_{p'}]a_{p'}[y_{p'}]a_{p}[y_p]x[y_p]\in\sigma$ by
hyperuniformity, we have $tNN'\in\tilsig$ for some $N'$ with
$\erase(N')=x[y_p]$. So $x=b[z]$ with $b\in\occ_{A\fl B}$, and we
choose $f(a_p)=b$. By the copycat property $z[y_p]=y_p$, so $z[i]=i$:
this means $f(a_1)[i]a_1[i]a_2[i]f(a_2)[i]\dots
a_{p}[i]f(a_p)[i]\in\sigma$.

\medskip

If
$p=p'+1$ with $p'$ even, one can do the same reasoning by using $\tau$
instead of $\sigma$.

\medskip

In the same way, one can associate a function $g$ with $\tau$ and
easily verify that $f\circ g$ is the identity on $\occ_B$ and $g\circ
f$ is the identity on $\occ_A$, so $f$ is a bijection. Moreover, by
construction, if $a\leq a'$ then $f(a)\leq f(a')$.

\bigskip

\subsection*{Construction of the typed copycat play}


To prove that $f$ satisfies the requirements of a Curry-isomorphism,
we will construct a play $s_p\in\tilde{\sigma}$ such that
$\erase(s_p)=t_p$, where
$t_p=\begin{cases}f(a_1)[y_1]a_1[y_1]a_2[y_2]f(a_2)[y_2]\dots
f(a_p)[y_p]a_{p}[y_{p}]&\text{if $p$ odd}\\
f(a_1)[y_1]a_1[y_1]a_2[y_2]f(a_2)[y_2]\dots
a_{p}[y_{p}]f(a_{p})[y_{p}]&\text{if $p$ even}\end{cases}$\\ for an
appropriate choice of the moves $y_i$. Moreover, one will have
$s_p=s_{p-1}m_1[m_2]M'$ where $\AAA(m_1)=c\in\occ_{A\fl B}$ and
$\erase(m_2)=y_p$ uniquely determined by $D=\frac{m_1}{\LLL_{A\fl
B}(c)}$.

\medskip


In the plays $s_p$, we will use the games $(C_j)_{j\in\mathbb{N}}$
defined by: $C_1=\bot\times\bot$ and $C_{j+1}=C_j\times C_j$. Note
that each initial move of $C_j$ takes the form
$b_1(b_2(\dots(b_{j}(0))\dots))$, where each $b_i$ can be either $r$
or $l$. We call $r_j$ the initial move of $C_j$ where each $b_i$ is
equal to $r$. These games will be used in order to have ``fresh''
moves, \ie\ moves that cannot come from a game defined before $C_j$ is
played. In what follows, the integer $n_p$ is made to ensure that no
game defined before step $p$ can belong to $C_q$ for $q\geq p$.

We now build the triple $(s_p,y_p,n_p)$ inductively:
\begin{itemize}
\item If $p=1$, we define the typed move $M_1=m_1[m_2]$ such that:
  $\AAA(m_1)=f(a_1)$, the $d_1$ games played at the level of $m_1$ are
  $C_1,\dots,C_{d_1}$ and we choose $m_2=\sharp(f(a_1))$ if $\paux_{A\fl B}(f(a_1))$ is
  undefined, $m_2=r_j$ if $\frac{m_1}{\LLL_{A\fl B}(f(a_1))}=C_j$. As
  $M_1\in\mc{P}_{A\fl B}$ and $\tilde{\sigma}$ is a realization of
  $\sigma$, there exists $M'_1$ such that $M_1M'_1\in\tilsig$ and
  $\erase(M_1M'_1)=f(a_1)[y_1]a_1[y_1]$, where $y_1=\erase(m_2)$. We
  choose $s_1=M_1M'_1$. Let us define $N$ as the biggest number of
  tokens $r$ in any initial occurrence of a game $D$ defined at the
  level of
  $M'_1$. We choose $n_1=\textit{max}(d_1,N)+1$.
\item If $p=p'+1$ with $p'$ odd, we define the typed move $M_p=m_1[m_2]$
such that: $\erase(m_1)=f(a_p)$, the $d_p$ games defined at the level of $m_1$ are
$C_{n_{p'}},\dots,C_{n_{p'}+d_p}$ and $m_2$ is chosen as follows:
\begin{itemize}
\item if $\paux_{A\fl B}(f(a_p))$ is undefined, $m_2=\sharp(f(a_p))$ 
\item if $\paux_{A\fl B}(f(a_p))=\mathbf{O}$, $m_2=r_j$ if
  $\frac{m_1}{\LLL_{A\fl B}(f(a_p))}=C_j$
\item if $\paux_{A\fl B}(f(a_p))=\mathbf{P}$, let
  $D=\frac{m_1}{\LLL_{A\fl B}(f(a_p))}$. Either there exists
  $c\in\occ_D$ such that $\vdash c$ and $\LLL_D(c)\neq\dag$, and in
  this case one chooses $m_2=m'_2[m'_3]$ with $\AAA(m'_2)=c$,
  $\frac{m'_2}{\LLL_D(c)}=\bot\times\bot$ and $m'_3=l0$; we note
  $r_D=\erase(m_2)\ $\footnote{In this case, $m_2$ is precisely built in such a way
  that we cannot have $r_D=r_j$ for any $j$.}, and we require that $r_D$ is a function of
  $D$; or there exists no such
  $c$ and in this case we choose $m_2$ such that $\AAA(m_2)$ is one of
  the initial occurrences of $D$: we just require that this choice is
  a function of $D$, and note it $r_D$.
\end{itemize}
As $s_{p'}M_p\in\mc{P}_{A\fl B}$ and $\tilde{\sigma}$ is a realization
of $\sigma$, there exists $M'_p$ such that $s_{p'}M_pM'_p\in\tilsig$
and $\erase(s_{p'}M_pM'_p)=t_p$ if $y_p=\erase(m_2)$. We choose
$s_p=s_{p'}M_pM'_p$. Let us define $N$ as the biggest number of tokens
$r$ in any initial occurrence of a game $D$ defined at the level of $M'_p$. We
choose $n_p=\textit{max}(n_{p'}+d_p,N)+1$.
\item If $p=p'+1$ with $p'$ odd, we do the same choices as in the
preceding case, except that $f(a_p)$ must be replaced by $a_p$, and
conversely.
\end{itemize}

Suppose $\paux_{A\fl B}(a_p)$ is defined, then $\rref_{A\fl B}(a_p)=b$
is also defined. It is important for the next section of the proof to
understand the link between $b$ and the play $s_p$. First, note that
$b=a_i$ for some $1\leq i\leq p$; then, because of the definition of
the set $\RRR_{A\fl B}$ of hyperedges, we know that $a_i$ is the
minimal occurrence $c$ of $\occ_{A\fl B}$ such that $\LLL_{A\fl
B}(a_p)$ is a prefix of $c$. Hence, if $M_i$ (resp. $M_p$) is the move
in $s_p$ such that $\erase(M_i)=a_i[y_i]$
(resp. $\erase(M_p)=a_p[y_p]$) and if $D=\frac{M_p}{\LLL_{A\fl
B}(a_p)}$, then the game $D$ is played by $\paux_{A\fl B}(a_p)$ at the
level of $M_i$. So, in the construction of $s_p$, $D$ has been played
at step $i$.

\medskip

We also need to build a play $u_p\in\tilde{\tau}$ such that
$\erase(u_p)=v_p$, where
$\\v_p=\begin{cases}a_1[y_1']f(a_1)[y_1']f(a_2)[y_2']a_2[y_2']\dots
a_{p'}[y_{p'}]f(a_{p'})[y_{p'}]&\text{if $p$ odd}\\
a_1[y_1']f(a_1)[y_1']f(a_2)[y_2']a_2[y_2']\dots
f(a_{p})[y_{p}']a_{p}[y_{p}']&\text{if $p$ even}\end{cases}$
\quad for an appropriate choice of
the moves $y_i'$.\\
The procedure is similar (we just need to swap
$\sigma$ and $\tau$). Note that we do not have in general
$u_p=\breve{s}_p$, or even $\erase(u_p)=\breve{w}_p$ with
$w_p=\erase(s_p)$, because the untyped moves $y_i$ and $y'_i$ may
differ.

\bigskip

\subsection*{Curry-isomorphism}


We are now going to prove that the bijection $f$ satisfies each
requirement of a Curry-isomorphism.

\bigskip

We first prove that $\DD_B\circ f=\DD_A$: suppose $\DD_A(a_p)=X_i$,
then $s_p=s_{p-1}MM'$ with $\erase(M)=a_p[i]$ and
$\erase(M')=f(a_p)[i]$; likewise, $u_p=u_{p-1}NN'$ with
$\erase(N)=f(a_p)[i]$ and $\erase(M')=a_p[i]$. If $\paux_{A\fl
B}(f(a_p))=\OO$ then one should have $i=r_j$ for some $j$ by
construction of $s_p$, which is impossible. If $\paux_{A\fl
B}(f(a_p))=\PP$ then $\paux_{B\fl A}(f(a_p))=\OO$ and one should have
$i=r_j$ for some $j$ by construction of $u_p$, which is
impossible. Then $\paux_{A\fl B}(f(a_p))$ is not defined, and
$\sharp(f(a_p))=i$ which means $\DD_B(f(a_p))=X_i$. Similarly,
$\DD_B(f(a_p))=X_i$ implies $\DD_A(a_p)=X_i$ as well.

\bigskip

We then prove that $f(\mc{S}_A)=\mc{S}_B$: if $a_p\in S$ with
$(t,S)\in\RRR_A$ for some $t$, suppose $\LLL_{A\fl B}(f(a_p))=\dag$.
If $\paux_{A\fl B}(a_p)=\OO$ then $s_p=s_{p-1}MM'$ with
$\erase(M)=a_p[y_p]$ and $\erase(M')=f(a_p)[y_p]$, and one should have
$y_p=r_j$ for some $j$ by construction of $s_p$. But this is
impossible since $\LLL_{A\fl B}(f(a_p))=\dag$ implies
$\AAA(M')\in\occ_{A\fl B}$, so $y_p\in\mathbb{N}$. If $\paux_{A\fl
B}(a_p)=\PP$ then $\paux_{B\fl A}(a_p)=\OO$, $u_p=u_{p-1}NN'$ with
$\erase(N)=f(a_p)[y_p]$ and $\erase(N')=a_p[y_p]$ and one should have
$y_p=r_j$ for some $j$ by construction of $u_p$. But this is
impossible since $\LLL_{A\fl B}(f(a_p))=\dag$ implies
$\AAA(N)\in\occ_{A\fl B}$, so $y_p\in\mathbb{N}$.

\bigskip

Finally, we need to prove the following: for every $(t,S)\in\RRR_A$,
if there exists $c\in S$ such that $\lambda(c)\neq\lambda(t)$, then
$(f(t),f(S))\in\RRR_B$ (the reciprocal would be done similarly).  Let
us take $a_1,\dots,a_p$ the sequence of nodes such that: $\vdash a_1$,
$a_i\vdash a_{i+1}$ and $a_p=c$. We necessarily have $t=a_i$ for some
$i\leq p$.

First we prove that $\rref_B(f(a_p))=f(a_i)$: suppose that it is
false, then $\rref_B(f(a_p))=f(a_j)$ with $j\neq i$.
First take
$j<i$: if $\paux_{A\fl B}(a_p)=\mathbf{O}$, then $f(a_p)$ is an $\OO$-move on
$A\fl B$, so $s_p=SM_pM'_p$ where: $M_p=m_1[m_2]$, $\AAA(m_1)=f(a_p)$
and $\frac{m_1}{\LLL_{A\fl B}(f(a_p))}=D$ for some $D$
chosen at step $j$; and $M'_p=m'_1[m'_2]$, $\AAA(m'_1)=a_p$ and
$\frac{m'_1}{\LLL_{A\fl B}(a_p)}=C_{k'}$ for some $k'\geq n_{i-1}$. So
we should have $y_p=r_k$ to be the move we choose in $D$, which is
impossible by construction of $n_{i-1}$. If $\paux_{A\fl
  B}(a_p)=\mathbf{P}$, we simply note that $\paux_{B\fl
  A}(a_p)=\mathbf{O}$ and do the same reasoning with $u_p$ in $B\fl
A$.  In the case where $i<j$, the reasoning is similar: if $\paux_{A\fl
  B}(a_p)=\mathbf{P}$, then $s_p=SM_pM'_p$ where: $M_p=m_1[m_2]$,
$\AAA(m_1)=a_p$ and $\frac{m_1}{\LLL_{A\fl B}(a_p)}=D$ for some $D$
chosen at step $i$; and $M'_p=m'_1[m'_2]$, $\AAA(m'_1)=f(a_p)$ and
$\frac{m'_1}{\LLL_{A\fl B}(f(a_p))}=C_{k'}$ for some $k'\geq
n_{j-1}$. This leads to a contradiction. If $\paux_{A\fl
  B}(f(a_p))=\mathbf{P}$, we work on $B\fl A$.

Let us now have $b\in\friends_A(a_p)$, and suppose that
$f(b)\notin\friends_B{f(a_p)}$. By what has been proved before we know
that $\paux_{A\fl B}(f(b))$ is defined, but also that $\rref_B(f(b))$
has the same polarity as $\rref_A(b)$: indeed, if
$\paux_A(b)\neq\lambda(b)$ then $\rref_B(f(b))=f(\rref_A(b))$, so
$\lambda(\rref_B(f(b)))=\lambda(f(\rref_A(b)))=\lambda(\rref_A(b))$; similarly,
if $\paux_B(f(b))\neq\lambda(f(b))$ then
$\rref_A(b)=f^{-1}(\rref_B(f(b)))$, so
$\lambda(\rref_A(b))=\lambda(f^{-1}(\rref_B(f(b))))=\lambda(\rref_B(f(b)))$.
Finally, if $\paux_A(b)=\lambda(b)$ and $\paux_B(f(b))=\lambda(f(b))$
then $\paux_A(b)=\paux_B(f(b))$ because $b$ and $f(b)$ have the same
polarity. Then, in all cases, $\paux(b)=\paux(f(b))$.

We consider that $\paux_{A\fl B}(f(b))=\mathbf{O}$ (if not, one works
with $u_p$ on $B\fl A$), so $\paux_{A\fl B}(a_p)=\mathbf{P}$ and
$s_p=s_{p-1}m_1[m_2]m'_1[m'_2]$ with $\frac{m'_1}{\LLL_{A\fl
B}(f(a_p))}=C_k$ for some $k$. Let $D=\frac{m_1}{\LLL_{A\fl
B}(a_p)}$, we necessarily have that $y_p=r_k=\erase(r_D)$. But a
problem arises with $b$ and $f(b)$: as a first case, suppose that $b$
has the polarity $\PP$ in $A$. Then there is a play
$s'_{q}=s'_{q-1}M_1[M_2]M'_1[M'_2]$ in $\tilde{\sigma}$ constructed
the same way as $s_p$, such that
$\erase(s'_{q})=\erase(s'_{q-1})b[y'_{q}]f(b)[y'_{q}]$, and where
$\frac{M_1}{\LLL_{A\fl B}(b)}=D$ and $\frac{M'_1}{\LLL_{A\fl
B}(f(b))}=C_{k'}$ with $k'\neq k$. Then we should have
$y'_q=\erase(r_D)$ occurrence of $C_{k'}$, so $r_k=r_{k'}$ which is
impossible.

The second case is where $b$ has the polarity $\OO$ in
$A$. Then there is a play $s'_{q}=M_1[M_2]M'_1[M'_2]$ in
$\tilde{\sigma}$ constructed the same way as $s_p$, such that
$\erase(s'_{q})=s'f(b)[y'_{q}]b[y'_{q}]$, and where
$\frac{M_1}{\LLL_{A\fl B}(f(b))}=C_{k'}$ with $k'\neq k$ and
$\frac{M'_1}{\LLL_{A\fl B}(b)}=D$. Then we should have
$y'_q=r_{k'}=\erase(d')$ with $d'$ move in $D$. But in this case
$\AAA(d),\AAA(d')\in\occ_D$ (if not we have a token $l$ in $d$ or
$d'$), so $\mc{E}(d)=r_{k}$ and $\mc{E}(d')=r_{k'}$, hence $k=k'$
because $D$ is unambiguous. This is impossible.

$f(b)\in\friends_B(f(a_p))$ similarly implies $b\in\friends_A(a_p)$,
so $f(S)=\{b\mid\ s\in\friends_B(f(a_p))\}$.  This allows us to
conclude that $(f(t),f(S))\in\RRR_B$.
\cqfd

\bigskip
\bigskip

\end{document}